\documentclass[a4paper,10pt]{article}
\usepackage{jheppub}
\usepackage[utf8]{inputenc}
\usepackage[english]{babel}
\usepackage{verbatim}
\usepackage{amsfonts}
\usepackage{amsthm}
\usepackage{mathrsfs}
\usepackage{enumitem}
\usepackage{relsize}
\usepackage{exscale}
\usepackage{array}

\numberwithin{equation}{section}

\allowdisplaybreaks[4]
\notoc 

\newtheoremstyle{mysty}{}{}{}{}{\bfseries}{.}{ }{\thmname{#1}\thmnumber{ #2}\thmnote{ (#3)}}
\theoremstyle{mysty}

\newcommand{\MyCol}[3]{\begin{matrix}
#1 \\ #2 \\ #3                        
\end{matrix}}

\newcommand{\SixJ}[6]{\begin{Bmatrix}
#1 & #2 & #3 \\
#4 & #5 & #6 
\end{Bmatrix}}
\newcommand{\NineJ}[9]{\begin{Bmatrix}
#1 & #2 & #3 \\
#4 & #5 & #6 \\
#7 & #8 & #9
\end{Bmatrix}}
\newcommand{\FifteenJ}[5]{\begin{Bmatrix}
#1 & #2 & #3 & #4 & #5                        
\end{Bmatrix}}

\notoc

\begin{document}

\title{Renormalization of group field theories for quantum gravity: new computations and some suggestions}
\author[a,c]{Marco Finocchiaro} 
\author[b]{Daniele Oriti}
\affiliation[a]{Max Planck Institute for Gravitational Physics (Albert Einstein Institute), Am Muehlenberg 1, D-14476 Potsdam-Golm, Germany, EU}
\affiliation[b]{Arnold-Sommerfeld-Center for Theoretical Physics, Ludwig-Maximilians-Universit\"at, Theresienstrasse 37, D-80333 M\"unchen, Germany, EU}
\affiliation[c]{Institute for Physics, Humboldt-Universit\"at zu Berlin, Newtonstraße 15, 12489 Berlin, Germany, EU}
\emailAdd{marco.finocchiaro@aei.mpg.de}
\emailAdd{daniele.oriti@physik.lmu.de}

\date{\today}
\abstract{We discuss motivation and goals of renormalization analyses of group field theory models of simplicial 4d quantum gravity, and review briefly the status of this research area. We present some new computations of perturbative GFT (spin foam) amplitudes, concerning in particular their scaling behaviour, and the numerical techniques employed to obtain them. Finally, we suggest a number of research directions for further progress.}

\maketitle
\section{Introduction}
Group field theories (GFT) \cite{Oriti:2011jm, Krajewski:2012aw, Oriti:2014uga} are quantum field theories which aim at describing the fundamental quantum structures that constitute spacetime. They are quantum field theories  {\it of} spacetime, rather than {\it on} spacetime. They are defined on group manifolds (hence the name), with an associated phase space given by the cotangent bundle of the same group.
$$
\varphi \,:\, G^{\times d} \rightarrow\, \mathbb{C} \; .
$$
Equivalently, in conjugate variables, the basic field maps $d$ copies of the Lie algebra of the same group to the complex numbers, and should be understood in general as a non-commutative function (i.e. an element of a non-commutative algebra of functions), since the Lie algebra is in general a non-commutative manifold and this reflects on the algebra of functions defined on it. Depending on the specific model one is considering, various restriction can be imposed on the field, its domain, its target, and of course the choice of group manifold and \lq dimension\rq $d$ are also model-dependent. What is general, in all current GFT models, is that the basic quanta of the theory, corresponding to the basic field excitations, can be represented as abstract cells or polyhedra with the $d$ algebraic data forming the domain of the field associated to their (boundary) faces. When $d$ is chosen as the dimension of the spacetime to be reproduced in some approximation, the corresponding GFT quanta can be understood as $(d-1)$-simplices with algebraic labels on their $d$ $(d-2)$-dimensional faces. This is the case we restrict to in the following. The other feature which constitutes another defining aspect of the formalism is the peculiar combinatorial structure of field interactions. The dynamics of the theory, which dictates how the fundamental GFT quanta interact forming extended spacetime structure, is specified by an action, first, 

$$
S(\varphi,\varphi^*)\,=\,\frac{1}{2}\int[dg]\,\varphi^*(\tilde{g}_i)\,\mathcal{K}(\tilde{g}_i,g_i)\,\varphi(g_i)\; +\;\frac{\lambda}{n!}\int[dg_{ia}]\,\varphi(g_{i1})...\varphi(g_{in})\,\mathcal{V}(g_{ia}) \; +\; c.c.
$$

and then by its partition function (assumed here as being of statistical form):

$$
Z\,=\, \int \mathcal{D}\varphi\mathcal{D}\varphi^*\; e^{-\, S_\lambda(\varphi,\varphi^*)}\; .
$$

Beside a quadratic local term (defined by an integral kernel convoluted with the two fields), the GFT action is determined by interactions (just one in the above example) that possess a characteristic \lq combinatorial non-locality\rq in that the interaction kernels pair non-locally the field arguments ($d$ variables being contributed by each field entering the given interaction term). Interaction kernels of order $n$ can be associated with possible ways of gluing together $n$ $(d-1)$-simplices to form (the boundary of) a $d$-dimensional cell. The specific combinatorial patterns (i.e. the specific cells being associated to each fundamental interaction) and precise form of the interaction and kinetic kernels are part of the definition of each particular GFT model. However, from this generic aspect  of the formalism follows one key fact: GFT Feynman diagrams $\Gamma$ generated by the perturbative expansion of the GFT partition function, 
$$
Z\,=\, \int \mathcal{D}\varphi\mathcal{D}\varphi^*\; e^{-\, S_\lambda(\varphi,\varphi^*)}\;=\;\sum_\Gamma \frac{\lambda^{N_\Gamma}}{sym(\Gamma)}\,\mathcal{A}_\Gamma
$$
obtained gluing interaction vertices ($d$-cells) along their $(d-1)$-faces, are dual to $d$-dimensional cellular complexes, of arbitrary topology (since a priori there is no restriction on the allowed gluings). The Feynman amplitudes assign a probability amplitude for each such cellular complex, seen as an elementary interaction process of the fundamental GFT quanta.

The above can be taken as a sketchy definition of the TGFT formalism, but it is of course the specification of particular models which gives tentative physical meaning to it and this meaning will therefore change in different contexts. In particular, when the physical interpretation of a given model is grounded in its perturbative expansion, it will affect what we expect about the properties of the Feynman amplitudes $\mathcal{A}_\Gamma$. In this contribution, we focus on GFT models for quantum gravity and in particular on the class of models closely related to simplicial gravity path integrals, spin foam models and loop quantum gravity. For this class of models we discuss the key features of the Feynman amplitudes in the next subsection. In particular, beyond their interpretation, we will discuss at length whether we should expect them to be finite or divergent when dwelling into the issue of renormalization of the corresponding quantum gravity models.

\

Before doing so, let us spend some words to clarify the choice of a statistical form for the GFT partition function. The foundations of the GFT formalism have received some attention only recently, and much remains to be understood. Both statistical and complex weights for GFT fields summed over in the definition of the partition function can be found in the literature, but in fact most of the literature until recently has been focusing only on the perturbative expansion of the models, where the non-perturbative definition is less directly relevant, thus avoiding the need to investigate it. In particular, for the class of models to be analysed in the next subsection, the interpretation in terms of quantum sums over discrete geometries (and topologies) is guaranteed by the form of the Feynman amplitudes (which are either complex or real with oscillating behaviour, in line with their interpretation as defining a quantum discrete path integral for gravity on a lattice, as we are going to discuss). As for the non-perturbative definition of the GFT dynamics, instead, we have less guidelines, especially for the general definition of the formalism, before considering specific models. In the context of tensor models and tensorial group field theories treated as a generalization of usual local QFT framework, all non-perturbative analyses have simply assumed from the start a statistical definition, taken to be primary and in no special need of further justification; in particular, no attempt to understand it as the counterpart of an operatorial definition of the same model has been made. In non-quantum gravity applications, this is simply a choice of a \lq classical\rq statistical field theory context, usually adopted for purely mathematical reasons (better chance of making sense of the path integral). 

For quantum gravity applications, one may be interested also in a better conceptual foundation of the formalism. In this context, a complex weight involving the GFT action as a pure phase is of course a possibility, and maybe more in line with the intuitive idea of a quantum gravity path integral (even though the quantum gravity interpretation of the GFT field itself is not straightforward, while as we remarked already, this interpretation be consistently associated to the discrete structures appearing in the GFT perturbative expansion). However, we should remind ourselves that these quantum gravity models should not be expected to encode any global unitary evolution, as it is in fact true for any fundamental quantum gravity dynamics, and this removes one strong motivation to insist on this formulation of the path integral. Nor we have a complete derivation of the GFT path integral from a canonical quantum gravity dynamics or, alternatively, from some formal field theoretic gravitational path integral, that could dictate one choice over another. To date, the only tentative derivation of a GFT partition function \lq from first principles\rq , was given in \cite{Chirco:2018fns} where it was seen to arise from a {\it quantum} statistical definition of equilibrium for a system of quantized simplices (indeed, the basic quanta of such GFT models), under a requirement of maximization of entropy and a choice of macroscopic conditions to be imposed on average (this choice concurs to the specification of the resulting GFT action). If among these constraints one includes some appropriate counterpart of the Hamiltonian constraint of canonical gravity (adapted to the discrete setting), then the GFT partition function can be seen as a sort of grandcanonical partition function relaxing the imposition of such constraint (in the sense that configurations satisfying the constraint are assigned greater weight, but fluctuations off the constraint surface are allowed) as compared with a \lq microcanonical\rq ensemble in which only solutions of the constraint are allowed. From a canonical quantum gravity perspective the latter, more restricted case would correspond to a definition of the physical inner product between quantum gravity states (with appropriate insertions of observables inside the GFT partition function), while in general the GFT formalism deals then with a broader class of quantum amplitudes. This scenario was also anticipated (more formally) in \cite{Oriti:2013aqa}, discussing the relation between GFT and canonical loop quantum gravity. This perspective also resonates (with many details still to be clarified, though) with the presentation of GFT from a quantum gravity perspective in \cite{Freidel:2005qe}, where the \lq tree level\rq (thus dominant, from the perturbative point of view) GFT amplitudes were suggested to define the physical inner product of canonical quantum gravity, while the remaining GFT configurations were associated with topology changing processes (off-shell, from the canonical quantum gravity perspective).

\subsection{GFT, spin foams and other quantum gravity formalisms}
We recognize in this brief outline the straightforward generalization of how 2d surfaces are generated in the perturbative expansion of random matrix models. Indeed, GFTs can be seen as group-theoretic enrichment of random {\it tensor} models \cite{Gurau:2012hl, Rivasseau:2013uca, Gurau:2019qag}, to which they reduce if the Lie group domain is replaced by any finite set of $N$ elements. The Feynman amplitudes become purely combinatorial, but the type of diagrams remains the same. Seen as tensors, GFT fields admit a natural action of unitary (and orthogonal) groups on their arguments. If one requires GFT interactions to be invariant under such unitary transformations, they can be fully classified, and we speak of {\it tensorial} GFT models. Most of the literature on GFT renormalization \cite{Carrozza:2013mna, Carrozza:2016vsq} concerns these tensorial GFT models. More generally, focusing on tensorial aspects of GFTs allows to gain a greater control over the combinatorial structures of their states, diagrams and amplitudes and many of the results obtained in the simpler context of tensor models apply also to GFTs: the use of colors to encode the topology of Feynman diagrams, the large-N expansion, double scaling limits, universality results etc. The first two, in particular, are crucial for GFT renormalization.

In this contribution, we focus on GFT models which are \lq quantum geometric\rq : their fundamental quanta are quantised tetrahedra with a quantum geometry encoded in group-theoretic data. More precisely, the classical phase space of a single Lorentzian tetrahedron in 4d is chosen to be the cotangent bundle of 4 copies $SL(2,\mathbb{C})$, reduced by additional \lq geometricity\rq constraints, and in turn this can be mapped, under the same constraints, to the cotangent bundle of 4 copies of $SU(2)$. At this classical level, this map amounts simply to a change of variables between two alternative parametrizations of the same classical geometry of an individual tetrahedron. We will give more details on the simplicial geometry in the next subsection (see also \cite{Pereira:2010wzm}).  Appropriate gluing of five geometric tetrahedra on the boundary of a combinatorial 4-simplex can then be shown to provide a geometric characterization of the 4-simplex too, and of the whole simplicial complex obtained gluing geometric 4-simplices together. The same construction in the Riemannian case uses $Spin(4)$ instead of $SL(2,\mathbb{C})$. From the choice of classical phase space follows a choice of Hilbert space for individual quanta of the GFT model, given in one representation by $L^2(G^4)$ reduced by the quantum counterpart of the geometricity constraints, where $G$ is one of the chosen groups mentioned above. More precisely, the natural Hilbert space for a single tetrahedron in this class of models would be $L^2((SL(2,\mathbb{C})^4)$ restricted by the geometricity conditions dictated by the underlying classical simplicial geometric understanding of the GFT quanta (and amplitudes); this Hilbert space can be mapped, however, to the Hilbert space $L^(SU(2)^4)$, with the geometricity constraints encoded, in a model-dependent manner, in the definition of the map. In this latter case, the covariance properties of states and amplitudes under the action of the Lorentz group as well as some of their geometric features become \lq hidden\rq in the form of the kernels defining the GFT action or in the corresponding spin foam amplitudes, while boundary data (and GFT fields) only depend on $SU(2)$ data. Spin foam and GFT models defined using one or the other choice of fundamental Hilbert space are, in general, not equivalent, but the precise relation depends on the properties of the map being used (for example, its being isometric or not), but they would be equally justified from a simplicial geometric point of view. A definition of such map for the EPRL model has been given in \cite{Dupuis:2010jn}, together with an analysis of its properties, and a generalised definition of such map and analysis of its properties, valid for the whole class of models we deal with here, can be found in \cite{FJO}. The complete GFT Hilbert space is the corresponding Fock space built on this single-quantum Hilbert space. We will give a few more details in the next subsection.

For this class of geometric models, the GFT formalism benefits from direct links to other modern quantum gravity approaches, which can, viceversa, benefit from GFT tools and results. 

First, of all, when $SU(2)$ is used, the Hilbert space of a single GFT quantum is the same as that of a loop quantum gravity (LQG) \cite{Bodendorfer:2016uat} 4-valent spin network vertex. Generic GFT states, organized in a Fock space, will be populated by many such vertices and they will include, in particular, states corresponding to spin networks associated to closed graphs and gauge-invariant cylindrical functions for the same graphs. In fact, the LQG Hilbert space associated to any graph can be shown to be faithfully embedded in the GFT Fock space. The theories however differ in the way these graph-based Hilbert spaces are related, more precisely, in the scalar products between states associated to different graphs. Still, the correspondence, which can be extended to observables and quantum dynamics, allows to see GFTs as a 2nd quantized counterpart of LQG \cite{Oriti:2013aqa}.

Next. for this class of models, the GFT Feynman amplitudes take the form of (non-commutative) simplicial gravity path integrals \cite{Baratin:2011hp, Finocchiaro:2018hks}, when written in (non-commutative) Lie algebra variables, which encode the discrete metric. The group variables, on the other hand, are understood as encoding the discrete gravity connection. They correspond indeed to discretizations of a classical formulation of gravity as a topological BF theory with added geometricity constraints, on the simplicial complex dual to the GFT Feynman diagrams. The specific way in which the BF action is discretized depend on the quantization map applied to Lie algebra variables, and different models correspond to different strategies for the imposition of the constraints and path integral measures.

In fact, when the same Feynman amplitudes are recast as functions of group representations, using Peter-Weyl or Plancherel decomposition, they take the form of spin foam models \cite{Perez:2013uz}.  
Spin foam  models have been introduced as a covariant language for computing spin network dynamics,  so they can be understood as a covariant counterpart of canonical loop quantum gravity. A second perspective is to see spin foam amplitudes as a purely algebraic version of lattice gravity path integrals, or state sum models. In GFT, they arise as Feynman amplitudes. The correspondence is generic: for any given set of spin foam amplitudes associated to simplicial complexes (and admitting a local decomposition with respect to the complex), one can find a GFT action such that the perturbative expansion of the quantum partition function will produce the given amplitudes as Feynman amplitudes (and viceversa, any GFT action corresponds to a set of spin foam amplitudes). A complete definition of a spin foam model requires a prescription for the amplitudes to be associated to all possible cellular complexes (in some specified class) and an organization principle for them, i.e. one way of comparing, composing or selecting them, to obtain a single number for any observable one wants to compute. The GFT embedding provide one such clear organizing principle, by summing them in a QFT perturbative expansion. In addition, it provides a whole set of QFT tools that can be applied to study their mathematical foundations as well as for extracting physics. GFT renormalization can be seen, indeed, from this spin foam perspective.

\subsection{Simplicial GFT models for 4d quantum gravity}
The starting point for the construction of simplicial GFT (and spin foam) models of 4d quantum gravity is the quantum geometry of a single tetrahedron in 4d \cite{Dupuis:2012yw}. 

The quantum geometry of this basic building block, and the extended structured built from it, can be described in various parametrizations \cite{Baez:1999tk, Freidel:2010aq}, and a number of generalizations can also be defined \cite{Freidel:2018pvm, Freidel:2019ees} and imported in the GFT framework. Classically, one can use two equivalent characterization of a tetrahedral geometry, leading immediately to an algebraic translation. First, one can start with assigning four vectors $b_i^I\in\mathbb{R}^{3,1}$ to the four faces of the tetrahedron, forced to lie all in the same spacelike hypersurface with timelike normal $V$ (thus satisfying $b_i \cdot V = 0$), and thinking of them as normal to the same faces, with their modulus identified with their area, $b_i^I = A_i n_i^I$ (with $|n| =1$). The vectors are also forced to close to form the closed boundary of the tetrahedron, i.e. $\sum_i b_i = 0$. The vectors $b_i$, due to the constraints they satisfy, are actually elements of the vector space $\mathbb{R}^3$ which can be identified with the Lie algebra $\mathfrak{su}(2)$, after  it has been endowed with the corresponding Lie bracket. The resulting space $\mathfrak{su}(2)^{\times 4}$ is then the space of geometries for a single tetrahedron. It can also be seen as the cotangent space of the phase space $(\mathcal{T}^*SU(2))^{\times 4}$ which is then the phase space of a classical tetrahedron, purely expressed in terms of group-theoretic, algebraic data. The conjugate variables in $SU(2)^{\times 4}$ have the interpretation of parallel transports of a discrete connection along elementary paths from (the (bari)center of) the tetrahedron to the ((bari)center of its) boundary faces.  The dual graph made of these paths becomes the graph associated to a single spin network vertex (with four outgoing \lq open links\rq). In group representation, the corresponding Hilbert space is thus $L^2( SU(2)^4)$ (with Haar measure). 

An equivalent encoding of the classical geometry of a single tetrahedron uses directly the variables of discretized topological BF theory. All geometric quantities of a single tetrahedron can be computed starting from four bivectors $B_i^{IJ} \in \wedge^2\mathbb{R}^{3,1} \simeq \mathfrak{sl}(2,\mathbb{C})$ which close $\sum_i B_i^{IJ} = 0$ and satisfy the simplicity constraints $V_I (*B_i)^{IJ} = 0$ ($*$ is the hodge dual), with respect to the same timelike normal vector $V$. The phase space of a single tetrahedron can be taken to be the cotangent bundle $\mathcal{T}^*SL(2,\mathbb{C})^4$ and the Hilbert space to be $L^2(SL(2,\mathbb{C})^4)$. See also \cite{Pereira:2010wzm} for more details. This second construction can be indeed seen as the discrete (and then, quantum) counterpart of the formulation of continuum General Relativity as a constrained BF theory (a topological field theory) in 4 spacetime dimensions. This amounts to adding suitable constraints, called \lq simplicity constraints\rq, to the BF action, resulting in the B field of the topological theory being equivalent to a tetrad field, in such a way that the insertion of the general solution of these constraints in the BF action gives the Palatini formulation of classical continuum gravity in terms of tetrad and connection fields, in turn equivalent, at the classical level, to the metric formulation (modulo subtleties concerning degenerate geometries). The \lq geometricity\rq constraints we discussed above correspond to the combination of the (discrete counterpart of the) simplicity constraints and the gauge invariance constraints. For more details on the continuum formulation, see the cited references.
The two geometric descriptions can be mapped into each other, as we have mentioned already. The simplicity constraints can be seen also, in fact, as determining such map \cite{Dupuis:2010jn, FJO}. Since the Hilbert spaces indicated above admit a basis labeled by group representations, this correspondence can be seen also at that level, i.e. as specifying how the relevant representations of $SL(2,\mathbb{C})$ should be decomposed in $SU(2)$ representations, if they have to be understood as encoding the quantum geometry of a tetrahedron. Such representation labels are the variables in which spin foam amplitudes are expressed. Different spin foam (and GFT) models for 4d quantum gravity are specified (among other things) by the way the impose the simplicity constraints at the quantum level, and thus by the specific map between $SL(2,\mathbb{C}$ and $SU(2)$ entering their amplitudes, if used.  In the Riemannian case, which will be our focus in this contribution, all the above applies, with $Spin(4)$ replacing $SL(2,\mathbb{C})$.

A spin foam amplitude, that is a GFT Feynman amplitude written in representation variables, will be assigned to any given simplicial complex, dual to a GFT Feynman diagram. The basic building block is an assignment of a quantum amplitude to each 4-simplex, i.e. a \lq vertex\rq of the spin foam complex given by the GFT Feynman diagram, with this amplitude function of the algebraic data associated to the five tetrahedra on its boundary. These boundary data can be written as $SU(2)$ or $SL(2,\mathbb{C})$ data, using the mentioned map, and the vertex amplitude can be written as a function of both, featuring then the coefficients of the map, if used, and the geometricity constraints, in its expression. Thus the vertex amplitude will be a function of $SU(2)$ and $SL(2,\mathbb{C})$ representations associated to the triangles of the 4-simplex (faces of the dual complex), and intertwiners of both groups associated to the tetrahedra, following the imposition of the closure conditions (equivalent to gauge invariance with respect to both groups). The data not used as boundary data are then summed over independently in each vertex amplitude. The spin foam amplitude associated to the whole simplicial complex can then be obtained by gluing together the amplitudes associated to its 4-simplices, with the gluing amounting to matching first and then tracing over the data associated to the tetrahedron shared by each pair of 4-simplices, possibly weighted by an additional gluing kernel. In the GFT context, the vertex amplitude and the gluing kernel are nothing else than the interaction kernel and the propagator (inverse of the kinetic kernel) defining the GFT action. The correspondence between GFT amplitudes and spin foam models, which could be motivated and defined independently, is thus very general \cite{Reisenberger:2000zc}. 

One last comment about the discrete geometry of these models. The construction sketched above, at the classical level, leads to a full characterization of the discrete geometry of the 4d simplicial complex (to which the spin foam amplitudes are associated), equivalent to the more standard characterization in terms of edge lengths, as used in Regge calculus, even though it uses a different set of classical variables (it corresponds, indeed, to a formulation of classical simplicial geometry in terms of the discrete counterpart of the variables of BF theory, suitably constrained, or to so-called \lq area-angle\rq Regge calculus \cite{Dittrich:2008va}). The translation of the same characterization at the quantum level, and in particular the correct imposition of the geometricity constraints on quantum states and amplitudes, is the crucial point for ensuring the correctness of the model from a discrete geometric point of view, and it is still subject to debate in the literature. In particular, one would expect to find back the Regge action for metric (edge length) variables, or an equivalent classical reformulation, in the semi-classical expansion of the spin foam amplitudes for a generic simplicial complex (or of the corresponding simplicial path integral). Many results are available (see the cited references) on this issue for a single 4-simplex and for extremely simple complexes, at least for the EPRL model, but the results are mixed, and the situation is especially unclear for larger complexes. 

\

The general formula for the spin foam amplitudes, for all the models in this class, in the Riemannian setting, for given cellular complex $\mathfrak{m}$ dual to the Feynman diagram $\Gamma$ is the following:

\begin{align}
\mathcal{A}^{\beta}(\mathfrak{m}) = \sum_{J_{f}j_{ef}}\sum_{I_{ve}i_{e}}\prod_{f\in\mathcal{F}_{\mathfrak{m}}}d_{J_{f}}\prod_{e\in f}d_{j_{ef}}
\prod_{v\in\mathcal{V}_{\mathfrak{m}}}\{15J_{f}\}_{v}\prod_{e\in\mathcal{E}_{\mathfrak{m}}}
d_{I_{ve}}\sqrt{d_{i_{e}}}\,f^{i_{e},l/2}_{I_{ve}}(J_{f},j_{ef},k_{e},\beta)
\end{align}
The coefficients $f$ are matrix elements of the map between Spin$(4)$ and SU$(2)$ intertwiner spaces:
\begin{align}
&f^{i,l}_{I}(J_{p}, j_{p},k,\beta) = \sum_{M_{p}m_{p}}(\mathcal{I})^{J_{1}J_{2}J_{3}J_{4}I}_{M_{1}M_{2}M_{3}M_{4}}\bigg[\prod_{p=1}^{4}C^{j^{-}_{p}j^{-}_{p}j_{p}}_{m^{-}_{p}m^{-}_{p}m_{p}}(k)w^{l}(J_{p}, j_{p}, \beta)\bigg](\mathcal{I})^{j_{1}j_{2}j_{3}j_{4}i}_{m_{1}m_{2}m_{3}m_{4}} \label{FCI}
\end{align}
where we have a Spin$(4)$ representation $J_{f}$ labelling each face, a pair of Spin$(4)$ four-valent intertwiners $I_{ve},\,I_{v'e}$ for every edge and an SO$(3)$ spin $j_{ef}$ for each edge in a given face, while $C$ is the 3j-symbol, and the 15j-symbol is the one of the first type.
We have indicated with $w$ the function of group representations that characterizes the implementation of the simplicity constraints defining each model, depending on the representations of $Spin(4)$ and $SU(2)$ labelling each face of the complex. This depends also on the Immirzi parameter $\gamma$, through the combination $\beta = \frac{\gamma -1}{\gamma + 1}$.

The amplitude's formula can be rewritten in terms of propagators as follows:
\begin{align}
\mathcal{A}^{\beta}(\mathcal{G}) = \sum_{J_{f}}\sum_{I_{ve}}\prod_{f\in\mathcal{F}_{\mathcal{G}}}d_{J_{f}}\prod_{e\in\mathcal{E}_{\mathcal{G}}|f_{k}\supset e}\sqrt{d_{I_{ve}}d_{I_{v'e}}}\mathcal{K}^{J_{f_{1}}J_{f_{2}}J_{f_{3}}J_{f_{4}}}(I_{ve},I_{v'e},l,\beta)\prod_{v\in\mathcal{V}_{\mathcal{G}}}\{15J_{f}\}_{v}
\end{align}
where the propagator $\mathcal{K}$ is given by:
\begin{align}
&\mathcal{K}^{J_{1}J_{2}J_{3}J_{4}}(I,I',l,\beta) = \sum_{j_{1},\dots,j_{4}}\prod_{p=1}^{4}d_{j_{p}}\sum_{i}\sqrt{d_{I}d_{I'}}d_{i}\,f^{i,l/2}_{I}(J_{1},\dots,J_{4},j_{1},\dots,j_{4},\beta)
f^{i,l/2}_{I'}(J_{1},\dots,J_{4},j_{1},\dots,j_{4},\beta) \nonumber \\
&= \sum_{j_{1},\dots,j_{4}}\sum_{i}\sqrt{d_{I}d_{I'}}d_{i}
\prod_{p=1}^{4}d_{j_{p}}w^{l}(J_{p}, j_{p}, \beta)
\NineJ{j^{-}_{1}}{i^{-}}{j^{-}_{2}}{j^{+}_{1}}{i^{+}}{j^{+}_{2}}{j_{1}}{i}{j_{2}}
\NineJ{j^{-}_{3}}{i^{-}}{j^{-}_{4}}{j^{+}_{3}}{i^{+}}{j^{+}_{4}}{j_{3}}{i}{j_{4}}
\NineJ{j^{-}_{1}}{i'^{-}}{j^{-}_{2}}{j^{+}_{1}}{i'^{+}}{j^{+}_{2}}{j_{1}}{i}{j_{2}}
\NineJ{j^{-}_{3}}{i'^{-}}{j^{-}_{4}}{j^{+}_{3}}{i'^{+}}{j^{+}_{4}}{j_{3}}{i}{j_{4}}
\end{align}

\
In order to derive the master Integral expression for a GFT Feynman $\mathcal{G}$ we have to write down the corresponding (regularized) full amplitude $\mathcal{A}(\mathcal{G},\beta,\Lambda,\vec{J}_{\mathrm{ext}})$ and then set to zero all the spin associated to the external and contractible internal faces. We are going to give one concrete example of this procedure in the next section, when studying the scaling of the corresponding amplitude.
\

The models we will deal with in the following are the EPRL model \cite{Perez:2013uz} and the Duflo BO model \cite{Finocchiaro:2018hks}, whose defining maps are:
\begin{align}
&w_{EPRL}(j^{-},j^{+},j,\beta) = \delta_{j^{-}\,|\beta|j^{+}}\delta_{j\,(1+|\beta|)j^{+}} \quad \beta<0 \qquad w_{EPRL}(j^{-},j^{+},j,\beta) = \delta_{j^{-}\,|\beta|j^{+}}
\delta_{j\,(1-|\beta|)j^{+}} \quad \beta\geq 0 \label{WEPRL} \\
&w_{Duflo}(j^{-}, j^{+}, j, \beta) = \frac{(-1)^{j^{-} + j^{+} + j}}{\pi\,\sqrt{(2j^{-}+1)(2j^{+}+1)}}\,\sum_{a=0}^{\lambda}(\mathrm{Sign}(\beta))^{a}\SixJ{a}{j^{-}}{j^{-}}{j}{j^{+}}{j^{+}}\mathcal{T}_{a}^{j^{-}j^{+}}(\lvert\beta\rvert) \label{WDuflo}
\end{align} 
where the $T$ function is given by:
\begin{equation}
\mathcal{T}^{j^{-}j^{+}}_{a}(|\beta|) = (-1)^{a}(2a+1)\mathlarger{\int}_{0}^{2\pi}d\psi\,\frac{1}{|\beta|}\sin\frac{\psi}{2}\sin\frac{|\beta|\psi}{2}
\chi^{j^{-}}_{a}(\psi)\chi^{j^{+}}_{a}(|\beta|\psi) \qquad \lambda = 2\,\mathrm{Min}(j^{-},j^{+})
\end{equation}
but it can also be given an expression purely in terms of representation labels. See \cite{Finocchiaro:2018hks} for more details.

We note that the relative simplicity/complexity of these two models is highly dependent on the basis in which they are expressed, with the flux representation switching such relative complexity with respect to the spin representation given above.

We also point out that other spin foam models, obtained from alternative strategies of imposition of the same geometricity constraints and thus also belonging to the same general class we are considering, can be cast in principle in the same general form, and studied by the same method we will illustrate in the following. Beside difficulties, for some of them, in achieving an explicit and manageable  expression for their corresponding $w$ coefficients in representation variables, that makes the analysis more cumbersome, it would indeed be very interesting to perform the same scaling analysis of amplitudes and compare with our results.

More details about the construction of spin foam amplitudes, as well as all the ingredients we mentioned as entering in such construction, in a language well adapted to their GFT embedding, can be found in \cite{Finocchiaro:2018hks}.

\section{Renormalization of group field theories for 4d quantum gravity}
Let us now discuss motivation and current status of renormalization of simplicial GFT (and spin foam) models for quantum gravity. 

Beyond the connection to spin foam models and simplicial gravity path integrals, the general strategy for renormalization of GFT models \cite{Carrozza:2013mna, Carrozza:2016vsq, Baloitcha:2020idd} is to treat them as ordinary QFTs defined on a Lie group manifold, thus using the group structures (topology, Killing forms, etc) to define \lq scales\rq and mode integration. A natural notion of scale, to be used to label the RG flow, is provided by group representations, which index the spectrum of differential operators on the group, e.g. the Laplace-Beltrami operator, in turn often used to define the propagator of GFT models. Cut-offs imposed as part of a renormalization group scheme are then imposed on representation labels; for example, in the case of $SU(2)$ cutting off the spectrum of the Laplacian operator means imposing the bound $\sum_{i=1}^d j_i(j_i + 1) \leq \Lambda^2$, for some real (large) number $\Lambda$. This fits well with the fact that divergences in spin foam amplitudes mostly come from the large representations regime.
Still, a lot of non-trivial work (beside computational challenges) is needed to adapt for GFTs, whose Feynman diagrams are not graphs but cellular complexes, standard QFT notions, noticing also that any procedure for the contraction of divergent subgraphs of perturbative GFTs has the meaning, from the point of view of the simplicial gravity path integral or spin foam model corresponding to the Feynman amplitudes of the same, of a coarse graining scheme of the corresponding lattice theory. 

For a proper renormalization group scheme, however, two more ingredients are needed: control over the theory space corresponding to a given GFT model, i.e. the space of allowed interactions; a detailed characterization of the combinatorics of (the cellular complexes dual to) GFT Feynman diagrams. On neither of these two points much is known for simplicial 4d gravity models. As a result, most work in the context of GFT renormalization has been done focusing on tensorial GFT models, where the above limitations are not present. 

Before discussing the goals of GFT renormalization, we spend a few words of caution concerning the physical interpretation of the renormalization group scheme and derived flows. With scales associated to group representation labels, the natural cutoffs entering as UV cutoffs are for large representations. The associated RG then flows from large to small representations (from UV to IR). In LQG and simplicial quantum geometry representation labels identify eigenvalues of geometric operators (e.g. triangle areas or tetrahedral volumes). Large representation labels correspond to large values of such geometric quantities. Thus we have an apparent inversion of roles here, with large distances/volumes playing the role of UV scales in GFT. Caution however should be exercised. In both LQG and simplicial geometry, we know an area of a surface, say, to result roughly speaking from the sum of the individual areas of all elementary surfaces forming the one under consideration, so that one has $A \approx <j> <N>$ with $<j>$ the average area contribution and $<N>$  the (average) number of contributions. Moreover, experience from classical Regge calculus and other simplicial gravity formalisms, leads us to expect continuum geometry to be reproduced when the number of elementary excitations contributing to a given continuum geometric quantity is very large, with each contribution smallish (but allowed to be orders of magnitude above Planck size). On the same basis, we expect continuum geometry, and with it any notion of large or small areas, volumes, distances etc, to be the result of coarse graining microscopic, fundamental degrees of freedom like the ones we deal with in the fundamental GFT (or spin foam) formalism. We would better refrain, then, from interpreting simplicial observables directly as geometric, in the sense we attribute to continuum spacetime geometry and physics. Finally, one more alert comes from recalling that, in GFT, the simplicial geometric observables and excitations are the ones associated to the Fock representation of the theory, and probably this is does not correspond to a fully geometric phase in which continuum gravitational (thus, spatiotemporal) physics is to be found, being best adapted to perturbation theory around the fully degenerate (from the point of view of geometry) Fock vacuum.  

Also, let us comment on the importance of a better understanding of the symmetry properties of these 4d gravity models, both for the characterization of the corresponding theory space and for their relation to continuum gravity, which is of course crucially characterized by diffeomoprhism symmetry. The issue of symmetries in GFT models is very important but also very much open.  At the general level, we do not know much about symmetries of 4d GFT models, beyond the Lorentz invariance of the kinetic and interaction kernels and of the Feynman amplitudes (implemented as in usual lattice gauge theories, since the amplitudes are in fact lattice gauge theories for a (constrained) Lorentz connection). Because of the simplicity constraints and also of the simplicial combinatorics that characterize them, moreover, even the tensorial symmetry typical of tensor models is not present (or at least not manifest). Moreover, even for the few symmetries we know of, in other models, the analysis of their consequences, for example in terms of conservation laws, is complicated by the non-local nature of the GFT interaction (see the analysis \cite{Kegeles:2015oua, Kegeles:2016wfg}). Concerning diffeomorphisms, strictly speaking (being smooth transformations) they are not defined in a discrete context like that of GFT Feynman amplitudes, i.e. spin foam models and lattice gravity path integrals, and thus the question becomes whether we can identify some analogue of diffeomorphism symmetry that, in a continuum limit, could be then identified with the one characterizing GR. There are several analyses of such question for 3d (topological) models at the level of spin foam amplitudes \cite{Freidel:2002dw}, lattice gravity (see for example \cite{Dittrich:2008pw, Bahr:2009ku,Bahr:2009qc}) and corresponding GFT formulation \cite{Baratin:2011tg}, but nothing similar in the 4d gravity case (where the 4d counterpart of the symmetry identified in the 3d case is actually broken, at the discrete level \cite{Dittrich:2014rha}). When attempting a reconstruction of an effective dynamics of geometry in a continuum approximation, as done in the context of GFT cosmology, one has to proceed in terms of observables of the fundamental theory that have a chance to correspond to diffeomorphism invariant observables in GR, since all the structures of continuum GR on which diffeomorphisms act, e.g. manifold points, directions an coordinate functions, but also fields defined on the same manifold, are simply not present in the theory. 

\subsection{Quantum consistency and perturbative renormalization}
GFT models are first defined in perturbative expansion and it is in this perturbative formulation that spin foam amplitudes, and simplicial gravity path integrals, appear. The perturbative GFT amplitudes generically diverge and regularizations have to be imposed. It is this truncation that corresponds to working at a given \lq scale\rq. Is this definition of the quantum dynamics of GFT models consistent? is the spin foam description consistent? Here, consistency means first of all valid for all ranges of dynamical variables, under (controlled) removal of regulators. If not, the GFT model as defined in perturbative expansion, and thus the corresponding spin foam model (and simplicial path integral) cannot be trusted. In the GFT language, this is recognised immediately to be the issue of perturbative renormalizability of a given model.
We should only trust, from the spin foam or lattice gravity point of view, only GFT models that turn out to be (perturbatively) renormalizable.

We note in passing that there should be no requirement that the model is finite (in the sense of presenting no divergence even before any renormalization); first, we have no obvious reason to expect it, if the model contains an infinite number of degrees of freedom; second, renormalizable models are usually more interesting, as QFTs, than finite ones, since they have a non-trivial RG flow and new effective physics at each scale.

Let us clarify further what we mean, here, to avoid possible misunderstandings.  As a general point about field theories, we are saying that finiteness of the Feynman amplitudes associated to a given subset of diagrams, or even to all diagrams involved in a given "scattering process" is not so important, per se, and in fact not necessarily desirable. What is important is that the scattering amplitudes can be -made finite- by suitable renormalization procedure (at any order in perturbation theory), if originally divergent in terms of bare couplings, and after resumming all the diagrams involved in their computation. The final renormalized scattering (or transition) amplitudes are what is physically relevant. A theory that is instead simply finite in the sense of not requiring any renormalization, even if clearly easier to deal with, would be less interesting from a physical point of view because this finiteness would probably indicate that the quantum dynamics is not very rich and it does not change much across scales (i.e. when more of its quantum degrees of freedom are accounted for). The consequence would also be a less interesting phase diagram. This should explain our comments about finiteness of GFT Feynman amplitudes.
As for the finiteness of "quantum gravity scattering amplitudes", what we should expect or desire depends on how we interpret the terms. If we take a given GFT model to be a tentative definition of full quantum gravity, then for sure we should hope that its "transition or scattering amplitudes" be finite, in the end, i.e. after renormalization. If the relation between GFT and quantum GR is as in the first case discussed above, i.e. the two are "equivalent", then this also implies that the transition or scattering amplitudes of quantized GR will be finite, when properly defined and after renormalization (even if the renormalization procedure as well as the observables expressing the amplitudes may look very different in the two formulations). If the relation is as in the second case, and thus "quantum GR" is just an effective theory, then we do not have to expect that its transition or scattering amplitudes are finite tout court, but only within the domain of validity of the approximations or truncations leading to it within the fundamental theory.

Thus, the requirement of perturbative renormalizability is an important constraint, which helps removing from consideration inconsistent constructions. Here, the GFT embedding proves potentially very important also for spin foam models (and loop quantum gravity). All known GFT and spin foam models present several ambiguities, some intrinsic to any quantization procedure, others specific to simplicial GFT (and spin foam) models of quantum gravity. Requiring perturbative renormalizability means constraining such ambiguities. To name one, we have little constraints of the face amplitudes of spin foam models,, even though they can drastically affect the scaling behaviour of the GFT and spin foam amplitudes, to the point of allowing to achieve perturbative finiteness easily by simply fixing them to this end (which also shows why finiteness {\it per se} cannot be a goal, without a proper physical understanding), while perturbative renormalizability is a much trickier requirement. For example, see how simple modifications of the Barrett-Crane model (which is also the limit of both EPRL model and Duflo model for infinite Immirzi parameter) affect the resulting amplitudes \cite{DePietri:1999bx, Perez:2000fs}

Let us list some of them. A first one is combinatorial: why restricting to simplicial complexes? These are the ones for which we have a better understanding of the discrete geometry underlying our models, and in particular of the simplicity constraints that characterize them. But what other cellular complexes should be included in the theory for consistency, e.g. because corresponding to the counterterms required for taming the perturbative divergences? Others concern the underlying quantization and imposition of simplicity constraints. Being functions of the flux variables (which are non-commutative), they depend on which quantization map is chosen to quantize such variables. Different choices result in different discrete gravity actions and different simplicial path integral measures, thus different spin foam amplitudes. Also the very definition of the simplicity conditions as operator equations acting on quantum states depends on the chosen quantization map, from which follow thus different constraints on representation variables in the spin foam amplitudes. Further, the strategy by which simplicity constraints  are imposed produces in general different models or versions of the same type of models (this is apparent in the Riemannian case, while in the Lorentzian one we only have experience of different versions of the EPRL model). These and other ambiguities are discussed, e.g., in \cite{Finocchiaro:2018hks}. Using $SL(2,\mathbb{C})$ or $SU(2)$ data to label quantum states, which is another choice to make, also leads to potentially different models and amplitudes. Nor one should think that these ambiguities are an artefact of the GFT or spin foam formulation. They can be convincingly argued to be the counterpart of ambiguities in the definition of the canonical Hamiltonian constraint operator and, in a way, failing to fix (at least most of) them via renormalizability conditions would be the counterpart, at the background independent level, of the problem of non-renormalizability of perturbative quantum gravity on a given spacetime geometry \cite{Perez:2005fn}.
Perturbative GFT renormalizability is thus a crucial issue, also when one looks at it from the perspective of spin foam models, simplicial path integrals or canonical loop quantum gravity.
\\

So, where do we stand, on this important issue?
For simplicial GFT models of 4d quantum gravity the answer is, unfortunately, that we are only at the very beginning. The main reasons have been already mentioned. First, we do not know enough of their symmetries to characterize the relevant theory space. Second, the amplitudes for these models are very involved and technically challenging to compute, mostly due to the fact that the imposition of simplicity constraints makes them defined not simply on Lie group manifolds but on particular sub-manifolds of these (usually not even corresponding to homogeneous spaces). Third, dominant configurations (i.e. those giving the most divergent contribution to the amplitudes) are not just flat connections or similarly simple, but correspond to richer configurations from the point of view of simplicial geometry; possibly, they correspond to (or possibly include) the whole set of Regge geometries found as saddle-point configurations in the asymptotic analysis of spin foam amplitudes and corresponding simplicial path integrals. Therefore even power counting results are hard to obtain, and the brute force analysis of divergences is not advanced enough to indicate the needed counterterms, forming the theory space. If the theory space is hard to characterize also in the simpler 3d simplicial case (corresponding to topological BF theory), at least the amplitudes are manageable enough to obtain complete power counting theorems \cite{Bonzom:2011br}, identify some counterterms \cite{Geloun:2011cy} and nice finiteness results \cite{Geloun:2013zka}. 

So computational challenges are one big obstacle. It is on this aspect that we focus in the next section, presenting some new results in the Riemannian context.
These new results should be added to other ones we have on the calculation of radiative corrections and basic divergences of both Riemannian and Lorentzian simplicial spin foam models, and on explicit evaluations of their building blocks (mainly the vertex amplitudes). For a partial list, see \cite{Perini:2008pd}\cite{Geloun:2010vj}\cite{Bonzom:2013ofa}\cite{Riello:2013bzw}\cite{Chen:2016aag}\cite{Dona:2018pxq}\cite{Dona:2018nev}\cite{Gozzini:2019kui}\cite{Dona:2019dkf} and references therein. Future progress will build on these hard-won calculations. In turn these results build on the hard-won understanding, based on both analytical and numerical studies, regarding the asymptotic properties of SU$(2)$ recoupling coefficients and, more recently, of the SL$(2,\mathbb{C})$ recoupling invariants as well, following an extensive ongoing effort in the spin foam community to investigate the behaviour of the spin foam transition amplitudes for various models in the constrained BF theory class. A tentative (and incomplete) list of interesting references is \cite{Haggard:2009kv,2011PhRvA..83e2114Y,2011arXiv1104.3641Y,Bonzom:2011cy,Barrett:2009as,Barrett:2010ex,Barrett:2009gg,Johansson:2015cca,Speziale:2016axj,Dona:2017dvf,Dona:2020yao,Dona:2020xzv,Dona:2018nev,Dona:2018pxq,Dona:2019dkf,Gozzini:2019kui}.

For a comparison, one has to look at the amount of knowledge we have accumulated on tensorial (thus colored) GFT models \cite{Carrozza:2013mna, Carrozza:2016vsq, Baloitcha:2020idd}. Here we know several (classes of) models which are rigorously proven to be perturbative renormalizable, comprising both abelian and non-abelian models, on homogeneous spaces, with or without gauge invariance (closure condition), in different dimensions. Divergences are associated to bubbles, i.e. cells of the complex dual to the cellular complex associated to a GFT Feynman diagram, and typically the most divergent diagrams that form the relevant theory space of renormalizable theories are melonic ones, also singled out in tensor models. However, we also know example of TGFT models which are renormalizable outside the melonic truncation \cite{Carrozza:2017vkz}, and these examples may be relevant also for the case of simplicial GFT models, since the structure of their divergences presents some aspects of the simplicial case. 

\subsection{Continuum limit and non-perturbative renormalization}
GFT models of quantum gravity are bona fide QFTs, thus they possess infinite degrees of freedom, as we expect quantum gravity to do (at least thinking of it naively as a quantum theory of the gravitational field). Control over a very large number of degrees of freedom can only be achieved step by step, within some truncation scheme. With the inclusion of more and more degrees of freedom, we can expect a richer and richer set of new phenomena to be unraveled, simply because the physics of many (quantum, interacting) degrees of freedom is very different from that of few of them. In particular, we expect new phases to be revealed. Controlling the full quantum dynamics is controlling the continuum limit of GFT models, and this implies mapping out as best as we can the phase diagram of the same models. In practical terms, it means being able to evaluate the full GFT partition function, for given values of coupling constants. This is the problem of computing the full non-perturbative renormalization group flow of any given GFT model. 

Given the mentioned structural connections, understanding the non-perturbative renormalization of a quantum gravity GFT model implies controlling the continuum limit of the corresponding lattice gravity path integral and spin foam model, and the full quantum dynamics of the corresponding canonical loop quantum gravity formulation. 
The characterization of the continuum quantum gravity phase diagram and the identification of one phase where an effective general relativistic dynamics of spacetime can be extracted is in fact the key outstanding open issue in the field \cite{Oriti:2007qd, Oriti:2018tym, Dittrich:2014ala, Delcamp:2016dqo}.

This should already make clear why the precise relation between the (non-perturbative) renormalization of GFT models for 4d quantum gravity and the (non-perturbative) renormalization of continuum quantum GR treated as an ordinary field theory (as in the asymptotic safety approach) can only be envisaged in a very tentative manner. Let us give only some comment on our own tentative perspective on this. take a given GFT model that can be fully defined at the non-perturbative level, thus associated with a continuum phase diagram where RG flow trajectories are well-defined from the deep UV (in the GFT sense) to the full IR (still in the GFT sense), and thus accounting for all the (infinite) degrees of freedom of the model; to achieve this situation is the goal of non-perturbative GFT renormalization, as explained. A matching with GR requires that one can also compute, in the same model, observables which characterize fully a 4d geometry and that can be shown to satisfy the GR equations in a classical approximation. Now, we can envisage two possibilities. If the rewriting is, in the appropriate sense, exact, i.e. if one can in principle go from the GFT formulation of the theory to the geometric "quantum GR" one, in the same continuum limit, then the GFT model could be seen, in fact, as a definition of "quantum GR", without any change in dynamical degrees of freedom. In this case, one could expect that there exist a translation of the RG picture of the given GFT model into the one obtained by a non-perturbative RG treatment of GR, for example as provided (ideally) by the asymptotic safety scenario, and an isomorphism between their corresponding phase diagrams and RG trajectories. If the rewriting requires, instead, some truncation of the dynamical degrees of freedom of the GFT model, is valid only for a subset of the GFT observables, or some other drastic approximation to be valid, i.e. if "quantum GR" turns out to be only an effective, emergent description of some sector of the full quantum GFT, then the situation is different. In this case, we should not expect that the GFT phase diagram matches the GR one, and we can only expect that it will reproduce a portion of it, for scales and regime of couplings where the needed approximations and truncations hold. This regime will probably be the one corresponding to "low energies" from the standard GR and effective QFT perspective. Of course, all the above is very much tentative and it is hard to envisage the precise relation at the current stage of development of GFT as well as of "quantum GR", even though a number of features of GFT models (e.g. the fact that they include a sum over topologies and not just geometries, at least at the discrete level) would suggest that the second scenario is more likely.

\

Where do we stand, at the non-perturbative renormalization level? Beside work on the non-perturbative RG flow of tensor models \cite{Eichhorn:2018phj, Eichhorn:2019hsa}, a lot of activity has focused on the analysis of GFT models proper \cite{Carrozza:2016vsq, Baloitcha:2020idd}. 
Two main strategies have been followed. One is based on constructive methods, mostly focusing on the resummation of the perturbative series, e.g. showing Borel summability. The other is based on functional renormalization group analysis, either (mostly) based on the Wetterich-Morris equation for the effective action, or the Polchinski equation for n-point functions. For the same reasons that limited work on perturbative GFT renormalization, little is known about the general RG flow of simplicial GFT models of 4d quantum gravity. Simplicial GFT models in 3d have been shown to be Borel summable \cite{Freidel:2002tg, Magnen:2009at} and phase transitions for the GFT formulation of simplicial BF theory in any dimension has been shown to exist \cite{Baratin:2013rja}. But no similar analysis has been carried over to the 4d gravity case, where, as mentioned, we even lack perturbative indications. 

The tensorial GFT case, on the other hand, has been widely explored, mostly via functional renormalization techniques, with many results on a variety of models, again both abelian and non-abelian, with and without gauge symmetries, based on compact as well as non-compact Lie groups, in different dimensions. Concerning UV behaviour, asymptotic freedom is found in many examples and asymptotic safety is found in others \cite{Carrozza:2016tih}, in various truncations, and the perturbative results have been reproduced from a non-perturbative standpoint. More results on the relevance of truncation beyond the melonic sector have been found \cite{BenGeloun:2018ekd}, and the use of Ward identities for studying the RG flow have been explored \cite{Baloitcha:2020idd}. Concerning IR behaviour (i.e. the actual continuum limit), work is more limited (and more difficult) at the analytic level, but hints have been found, in various truncations and for various models, of a non-trivial phase diagram. In particular, hints of the existence of Wilson-Fisher fixed points (often found alongside asymptotic freedom in the UV) and of broken (or condensate) phases have been obtained \cite{Geloun:2016qyb, Carrozza:2017vkz}, indirectly supporting parallel work on the extraction of continuum gravitational physics from such condensate phases \cite{Oriti:2016qtz, Gielen:2016dss, Oriti:2016acw}. 

Even if it is unclear, at this stage, which of these results holds also in the simplicial 4d models, with their additional quantum geometric intricacies, all this work on tensorial GFT models has certainly led to a better understanding of GFT renormalization group schemes (and flows). This will certainly turn out to be useful also for the analysis of full-blown quantum gravity models.

\section{Bubble divergences and radiative corrections in GFT: some new results}
In this section we report on some recent results concerning the leading order radiative corrections to $N$-point functions ($N\leq 6$) for the \textit{Duflo model} and the EPRL model, whose amplitudes we have recalled above. We refer to the cited literature for more details on motivations, construction and features of these GFT (and spin foam) models. Also, we limit our presentation to a summary of results and procedures; a more detailed presentation with be left for a forthcoming publication.

\subsection[A warm-up example: the 2-point function of the Ooguri GFT model]{A warm-up example: the 2-point function of the Ooguri GFT model}
\begin{figure}[t]
\centering
\includegraphics[scale=1]{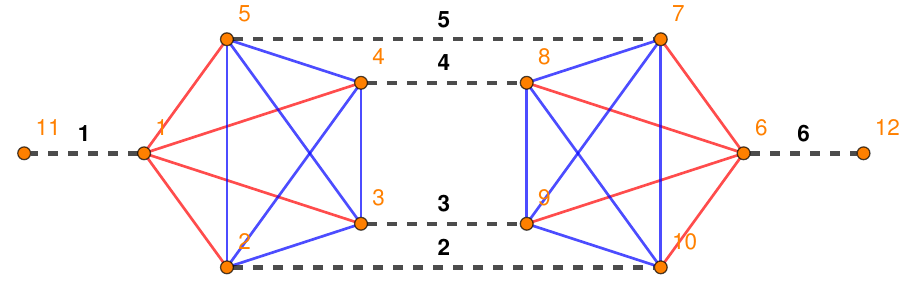}
\caption{The picture shows the LO (melonic) radiative correction to the simplicial Ooguri model two-point function $\mathcal{W}_{2}$.The dashed black lines represent propagators. The solid lines encode the internal structure of each simplicial GFT vertex. A face is an alternate sequences of dashed and solid lines. Blue lines are associated to internal faces while red lines belong to the external ones. The diagram has four external and six internal faces.}
\label{G221}
\end{figure}

As a warm-up, we recall the general procedure to compute the degree of divergences of GFT amplitudes both in the holonomy and spin formulation of GFT models, using the simpler case of the Ooguri GFT model for 4d topological BF theory with local $SU(2)$ invariance (that is, a simplicial GFT model on four copies of $SU(2)$ and with kinetic and interaction kernels made out of delta functions only).
\begin{description}[leftmargin=0pt]
\item {\it Analytical evaluation in group variables} - For models defined on the full group manifold or a corresponding homogeneous space (as it is the case in most tensorial GFT renormalization analyses), the evaluation is often conveniently done in the group representation. It can be carried out analytically, and it proceeds as follows. We compute first the bulk or amputated amplitude $\mathcal{A}_{bulk}$ by removing all the contributions from the external of the GFT diagram $\mathcal{G}$. This amounts to extracting only the dominant leading divergence of the amplitude (subleading divergences require a more refined procedure). Then we gauge fix the the holonomies on all the edges of a maximal rooted tree  of the graph $\mathcal{G}$. This reduces the evaluation to involve only a set of gauge-invariant variables. Next we drop the contribution from contractible internal faces\footnote{An internal face of $\mathcal{G}$ is contractible if it has at least one internal edge which is not shared with any other internal face.} (if any). The expression so obtained is the irreducible {\it Master Integral} $\mathscr{I}(\mathcal{G})$ associated to the amplitude, which can then regularized introducing appropriate cut-offs on the remaining integrals (e.g. by replacing the Dirac-delta functions with heat kernels). It is important to notice that different amplitudes contributing to different correlation functions might reduce to the evaluation of the same master integral. Last we evaluate the remaining integrals. This can be done analytically, exactly or approximately for example via saddle point methods (for example using the abelian asymptotic formula for the heat kernels), to find the Master integral scaling exponent $\omega(\mathcal{G})$, i.e. its superficial degree of divergence. 
\item {\it Numerical evaluation in the spin basis} - When the analytic evaluation in group variables is not possible, it is often more convenient to pass to the equivalent expression in terms of group representations (like the ones given above for 4d gravity models), and then proceed numerically, along similar steps as in group variables. First we compute the bulk or amputated amplitude by setting to zero all the spins labelling the external faces and using the appropriate identities for degenerate recoupling coefficients. Next we compute the Master Integral $\mathscr{I}(\mathcal{G})$, expressed in group representations, by setting to zero the spins labelling internal contractible faces. Its expression can be easily regularized by putting a uniform cutoff $\Lambda$ on all unbounded summations. Last we numerically evaluate the regularized master integral as a function of the cutoff. This can be done either using the full exact formula or using its approximate asymptotic formula (for large spins) obtained by uniform rescaling of all the spins. The amplitude degree of divergence $\omega$ can then be estimated by fitting the data. More precisely it is given by the angular coefficient of the linear best fit in a Log-Log data plot.    
\end{description}
Let us illustrate the general procedure with an example. We consider the leading order radiative (melonic) correction to the two-point function of this GFT model. The associated Feynman diagram is shown in Fig.\ref{G221}. The diagram has four external and six internal faces, none of which is contractible.
In the holonomy formulation, the amplitude can be written as follows:
\begin{equation}
\mathcal{A}_{\mathrm{Oo}}(\mathcal{G}_{melon}) = \mathlarger{\int}\bigg[\prod_{v\in\mathcal{G}}\prod_{e\ni v}dh_{ve}\bigg]\prod_{f\in\mathcal{F}_{\mathrm{cl}}}\delta\bigg[\prod_{e\in f}h_{ve}h_{v'e}^{-1}\bigg]\prod_{f\in\mathcal{F}_{\mathrm{ext}}}\delta\bigg[\prod_{e\in f}g_{f}h_{ve}h_{v'e}^{-1}\tilde{g}_{f}^{-1}\bigg] \quad .
\end{equation}
In this case, with or without gauge fixing, we can perform all the integrations exactly. Neglecting the contributions from the external faces, without a regularization we would find the divergent result:  
\begin{equation}
\mathcal{A}_{\mathrm{bulk}}(\mathcal{G}_{melon}) = \mathscr{I}_{\mathrm{Oo}}(\mathcal{G}_{melon}) = \delta(\mathbb{I})\delta(\mathbb{I})\delta(\mathbb{I}) \quad .
\end{equation}
The master integral can be regularized either via a sharp cut-off or by heat kernels:
\begin{align}
\delta(g)\longrightarrow \delta_{\Lambda}(g) = \sum_{j=0}^{\Lambda}(2j+1)\chi^{j}(g) \qquad
\delta(g)\longrightarrow K_{\alpha}(g) = \sum_{j=0}^{\infty}e^{-\alpha j(j+1)}(2j+1)\chi^{j}(g)
\end{align}
In both cases the amplitude's degree of divergence reads:
\begin{equation}
\mathscr{I}_{\mathrm{Oo}}(\mathcal{G}_{melon},\Lambda) \propto\Lambda^{9} \qquad \omega(\mathcal{G}_{melon}) = 9 \quad .
\end{equation}
The same result can be recovered by evaluating the amplitude in the spin basis. We have:
\begin{equation}
\mathcal{A}_{\mathrm{Oo}}(\mathcal{G}_{melon}) = \sum_{j_{f}|f\in\mathcal{F}_{\mathrm{cl}}}\sum_{i_{ve}}
\prod_{f\in\mathcal{F}_{\mathrm{cl}}}d_{j_{f}}
\prod_{e\in\mathcal{E}_{\mathrm{ext}}}(\mathcal{I})^{j_{f}i_{ve}}_{m_{ef}}\prod_{v\in\mathcal{G}}\{15j_{f}\}_{v} \quad .
\end{equation}
Since there are no contractible faces, by setting to zero all the spins labelling the external faces we immediately obtain the expression of the regularized master integral $\mathscr{I}_{Oo}$. Upon using the appropriate identify for the degenerate $15J$-symbol we find:
\begin{align}
&\mathcal{A}_{\mathrm{bulk}}(\mathcal{G}_{melon},\Lambda) = \mathscr{I}_{\mathrm{Oo}}(\mathcal{G}_{melon},\Lambda) = \sum_{j_{1}j_{3}j_{4}j_{7}j_{8}j_{9}}^{\Lambda}
d_{j_{1}}d_{j_{3}}d_{j_{4}}d_{j_{7}}d_{j_{8}}d_{j_{9}}\SixJ{j_{7}}{j_{3}}{j_{1}}{j_{9}}{j_{4}}{j_{8}}^{2} \\
&\mathscr{I}_{\mathrm{asy}}(\mathcal{G}_{melon},\Lambda) = \Lambda^{5}\sum_{j}^{\Lambda}d^{6}_{j}
\SixJ{j}{j}{j}{j}{j}{j}^{2} \approx \Lambda^{5}\sum_{j}^{\Lambda} j^{6}j^{-3}\approx \Lambda^{9} \quad .
\end{align}
The equation $3.6$ is the result of applying the identity $A.9$ to the Ooguri model's two point amplitude $\mathcal{A}_{\mathrm{Oo}}(\mathcal{G}_{221},\Lambda)$ (see the combinatorics of the diagram in Table 1) after setting to zero the spins labelling the external 
faces of the graph $\mathcal{G}_{221}$.

Thus the degree of divergence can be obtained by evaluating the master integral's exact formula, or  approximately from the above asymptotic formula, by combining the volume factor (replacing the redundant summations) and the face weights with the large-$j$ behaviour of the Wigner $6J$-symbol, obtaining:
\begin{equation}
\omega_{\mathrm{full}}(\mathcal{G}_{melon}) = 8.92 \qquad \omega_{\mathrm{asy}}(\mathcal{G}_{melon}) = 9
\end{equation}
in agreement with the analytical result obtained in the group formulation.

\subsection[Radiative corrections in simplicial GFT models for Quantum Gravity.]{Radiative corrections in simplicial GFT models for Quantum Gravity.}
We now report on some recent results concerning the leading order radiative corrections to $N$-point functions ($N\leq 6$) for the Duflo model and the EPRL model. 

We identify the relevant 1PI Feyman diagrams, compute the corresponding master integral formulae and use them to evaluate the master integrals' scaling (i.e. the diagrams' superficial degree of divergence) as a function of the cutoff. We also comment on the diagrams' combinatorial properties and on the structure of the corresponding counterterms. Finally we show how these results can be applied \lq beyond perturbation theory\rq to characterize to all orders the scaling of the necklace graphs (an important subclass of diagrams appearing in the radiative correction, also identified as the relevant graphs for renormalizability in the tensorial model of \cite{Carrozza:2017vkz}). 
We will derive first general formulae that applied to all models in the chosen class (again, simplicial models constructed from constraining those for topological BF theory) and then specialize to the models of interest by choosing the relevant form for the coefficients $w$, encoding the geometricity conditions characterizing them.  
\subsubsection[Leading order corrections to the N-point functions.]{Leading order corrections to the $N$-point functions}
The relevant 1PI GFT diagrams, appearing in the perturbative expansion of the $2$-point and $4$-point functions at the leading order in the GFT coupling constant $\lambda$, are shown in Tab.\,\ref{TabG2G4LONew} with the notation explained in the caption.
\begin{table}
\centering
\begin{tabular}[h]{m{0.3\textwidth}m{0.3\textwidth}}
\hline \\
\multicolumn{2}{c}{Leading order corrections to the $2$-point and $4$-point functions.}\\
\hline \\
\includegraphics[scale=0.55]{G221.pdf} & \includegraphics[scale=0.45]{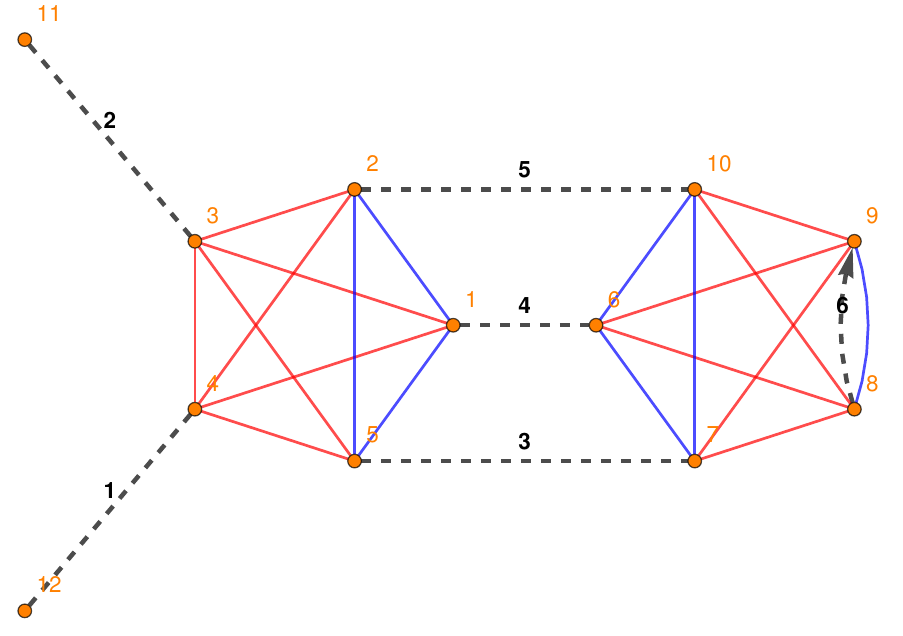} \\ \includegraphics[scale=0.55]{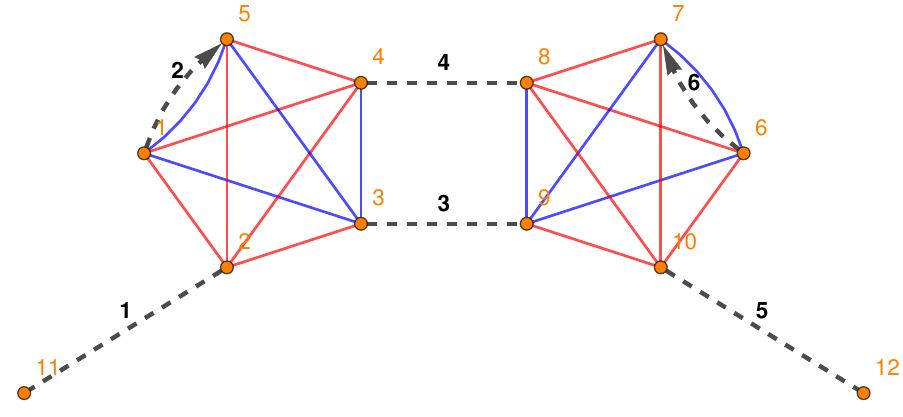} & \includegraphics[scale=0.5]{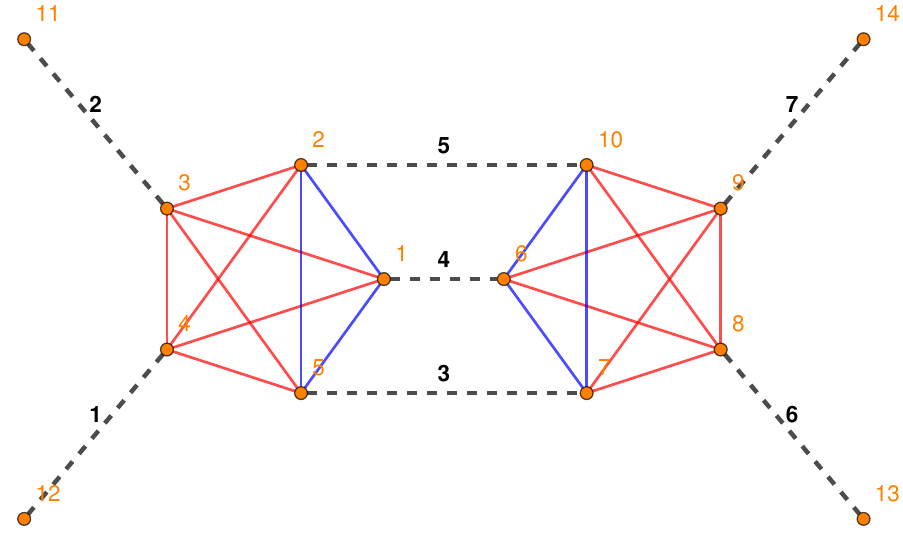} \\
\hline
\end{tabular}
\caption{The 1PI diagrams contributing to the LO expansion of the $2$-point and $4$-point functions $\mathcal{W}_{2}$ and $\mathcal{W}_{4}$. Each GFT Feynman diagram is labelled according to the number of external edges, the number of vertices and the position in the list. For example, the first graph on the left is called $\mathcal{G}_{221}$. The dashed lines represent the propagators. The solid lines denote the internal structure of each simplicial GFT vertex. Blue lines are associated to the internal faces, while red lines belong to the external ones. A face is a one-color alternating sequence of solid and dashed lines The labels for internal and external faces have not been shown.}
\label{TabG2G4LONew}
\end{table}
Let us make a few remarks before moving to the analysis of specific diagrams.
\begin{description}[leftmargin=0pt]
\item {\it Selected diagrams} - The diagrams showed in Fig.\,\ref{TabG2G4LONew} are the only potentially divergent 1PI GFT diagrams at the leading order. All the other leading order corrections to the $3-$, $5-$ and $6-$point functions, labelled $\mathcal{G}_{311}$, $\mathcal{G}_{511}$ and $\mathcal{G}_{621}$, are manifestly convergent. The diagram $\mathcal{G}_{511}$ has no internal faces, thus no potentially divergent summation over representations. The diagrams $\mathcal{G}_{311}$ and $\mathcal{G}_{621}$ have only one internal contractible face each. Therefore their corresponding Feynman amplitudes are again finite.
\item {\it $2-$point diagrams} - The diagram $\mathcal{G}_{221}$ is melonic and therefore also tracial. It has six internal faces and four external ones and it could be expected to be the most divergent LO contribution to the $2$-point function. Its dominant (or leading) divergence\footnote{The first melonic correction to the self energy might also have a subleading divergence as in the case of the Ooguri model and (Lorentzian) EPRL model. Such divergence is responsible for the wave function renormalization. This in turns requires a modification of the covariance with a second-order derivative term in order to account for the new counterterm.}, if any, can be subtracted by mass renormalization (as done in ordinary QFT). The associated irreducible master integral will be denoted as $\mathscr{I}_{221}$. The diagram $\mathcal{G}_{222}$ has four external faces and four internal faces; one of them, the tadpole face, is contractible. The diagram $\mathcal{G}_{223}$ has four internal faces and four external ones. Although the diagram $\mathcal{G}_{222}$ and $\mathcal{G}_{223}$ are not isomorphic, their Feynman amplitudes can be reduced to the evaluation of the $4-$point function's master integral $\mathscr{I}_{421}$. 
\item {\it The $4-$point diagram} - The diagram $\mathcal{G}_{421}$ is the first melonic correction to the $4$-point function. It has three internal faces (forming a bubble) and eight external ones. Its master integral $\mathscr{I}_{421}$, associated to the bubble subgraph, controls the UV scaling of the LO non-melonic $2$-point diagrams and of all the necklace diagrams contributing to the $N$-point functions with $N=4,5,6$.
\end{description}
To summarize: in order to determine the scaling behaviour and divergent structure of the leading corrections to the $2-$ and $4-$point functions we only need to study the independent master integrals $\mathscr{I}_{221}$ and $\mathscr{I}_{421}$, whose expression we will give and evaluate in the following.
\subsubsection[The 2-point function.]{The $2$-point function.}
To derive the expression of the master integral $\mathscr{I}_{221}$ for the leading (melonic) correction to the $2-$point function, as explained, we first write down the full regularized amplitude $\mathcal{A}(\mathcal{G}_{221},\beta,\Lambda,\vec{J}_{\mathrm{ext}})$ and then we set to zero all the spins associated to the external faces. Exploiting the identities (\ref{NineJA}, \ref{FifteenJD1}, \ref{FCII}) and a number of algebraic simplifications, we obtain:
\begin{align}
\mathscr{I}(\mathcal{G}_{221},l,\beta,\Lambda) &= \sum_{J_{f}\lvert f\in\mathcal{F}_{cl}}^{\Lambda}
d_{J_{1}}d_{J_{3}}d_{J_{4}}d_{J_{7}}d_{J_{8}}d_{J_{9}} 
\nonumber \\
&\times\mathcal{K}^{J_{3}J_{1}J_{7}}(l,\beta)
\mathcal{K}^{J_{4}J_{7}J_{8}}(l,\beta)
\mathcal{K}^{J_{8}J_{9}J_{3}}(l,\beta)
\mathcal{K}^{J_{9}J_{1}J_{4}}(l,\beta)
\SixJ{J_{7}}{J_{3}}{J_{1}}{J_{9}}{J_{4}}{J_{8}}^{2}
\label{I221New}
\end{align}
where the integer $l$ denotes the number of simplicity constraint insertions and the propagator\footnote{According to the diagram's connectivity, each internal edge of the graph $\mathcal{G}_{221}$ is shared by four faces (three internal and one external). Hence, after setting to zero the external spins, the internal propagators depend only on three variables.} $\mathcal{K}$ is given by the Eq.\,(\ref{KEff}). As anticipated, in the above formula, all the model-dependent features are encoded by the single-link fusion coefficients $w$ appearing in the expression $\mathcal{K}$, which we will refer to as the \lq propagator\rq ~in the following. Thus the above result is valid for \textit{any} simplicial GFT model for constrained BF theory. 

\

Before applying this formula to specific models, let us give a few more details on how it derived, the procedure being in fact the same for the other diagrams.

The melonic $2$-point diagram $\mathcal{G}_{221}$ has four external faces and six non-contractible internal faces. 
Furthermore each internal edge of the graph $G_{221}$ is shared by four faces (three internal and one external). The associated amplitude is given by:
\begin{align}
&\mathcal{A}(\mathcal{G}_{221},l,\beta,\Lambda) \varpropto \sum_{J_{f}\lvert f\in\mathcal{F}_{cl}}^{\Lambda}
\sum_{I_{ve}}\prod_{\mathrm{All}\,f}d_{J_{f}}\prod_{\mathrm{All}\,(ve)}\sqrt{d_{I_{ve}}}
\,\mathcal{K}^{J_{6}J_{1}J_{3}J_{7}}(I_{12},I_{22},l,\beta)
\mathcal{K}^{J_{7}J_{2}J_{4}J_{8}}(I_{13},I_{23},l,\beta) \nonumber \\ 
&\times
\mathcal{K}^{J_{8}J_{3}J_{5}J_{9}}(I_{14},I_{24},l,\beta)
\mathcal{K}^{J_{9}J_{4}J_{1}J_{10}}(I_{15},I_{25},l,\beta)
\FifteenJ{\MyCol{I_{11}}{J_{6}}{J_{1}}}{\MyCol{J_{2}}{J_{7}}{I_{12}}}{\MyCol{I_{13}}{J_{8}}{J_{3}}}{\MyCol{J_{4}}{J_{9}}{I_{14}}}{\MyCol{I_{15}}{J_{10}}{J_{5}}}
\FifteenJ{\MyCol{I_{26}}{J_{6}}{J_{1}}}{\MyCol{J_{2}}{J_{7}}{I_{22}}}{\MyCol{I_{23}}{J_{8}}{J_{3}}}{\MyCol{J_{4}}{J_{9}}{I_{24}}}{\MyCol{I_{25}}{J_{10}}{J_{5}}}
\end{align}
Hence after setting to zero the spins $J_{i} = (j^{-}_{i}, j^{+}_{i})$ labelling the external faces (in this case $J_{2},J_{5},J_{6},J_{10}$ according to the labelling conventions adopted in the paper), the four internal propagators will depend only on three spins. Thus in order to dervive the Master Integral expression $\mathscr{I}_{221}$ we first need to compute the formula for the degenerate propagator with one vanishing triad which in turns requires the identity for a \textit{type-A} NineJ symbols. Upon using the identity $A.1$ we obtain the degenerate propagator's formula $A.12$ which together with the $A.9$ leads us to the Master integral expression $3.9$. The same line of reasoning applies to the derivation of the Master Integral formula for any GFT graph with one and only one external leg for each simplicial vertex (like the melonic $2$-point function, upon setting the external spins to zero the internal propagators of these graphs will only depend on three spins).

\

For completeness, the full formula for the propagator is given by:
\begin{align}
&\mathcal{K}^{J_{1}J_{2}J_{3}J_{4}}(I,I',l,\beta) = \sum_{j_{1},\dots,j_{4}}\prod_{p=1}^{4}d_{j_{p}}\sum_{i}\sqrt{d_{I}d_{I'}}d_{i}\,f^{i,l/2}_{I}(J_{1},\dots,J_{4},j_{1},\dots,j_{4},\beta)
f^{i,l/2}_{I'}(J_{1},\dots,J_{4},j_{1},\dots,j_{4},\beta) \nonumber \\
&= \sum_{j_{1},\dots,j_{4}}\sum_{i}\sqrt{d_{I}d_{I'}}d_{i}
\prod_{p=1}^{4}d_{j_{p}}w^{l}(J_{p}, j_{p}, \beta)
\NineJ{j^{-}_{1}}{i^{-}}{j^{-}_{2}}{j^{+}_{1}}{i^{+}}{j^{+}_{2}}{j_{1}}{i}{j_{2}}
\NineJ{j^{-}_{3}}{i^{-}}{j^{-}_{4}}{j^{+}_{3}}{i^{+}}{j^{+}_{4}}{j_{3}}{i}{j_{4}}
\NineJ{j^{-}_{1}}{i'^{-}}{j^{-}_{2}}{j^{+}_{1}}{i'^{+}}{j^{+}_{2}}{j_{1}}{i}{j_{2}}
\NineJ{j^{-}_{3}}{i'^{-}}{j^{-}_{4}}{j^{+}_{3}}{i'^{+}}{j^{+}_{4}}{j_{3}}{i}{j_{4}}
\end{align}
The formula for the degenerate propagator (with one vanishing triad) can be written as follows:
\begin{equation}
\mathcal{K}^{J_{1}J_{2}J_{3}\,0}(I,I',l,\beta)\equiv \mathcal{K}^{J_{1}J_{2}J_{3}}(I,I',l,\beta) = \delta_{IJ_{3}}\delta_{I'J_{3}}\sum_{j_{1}j_{2}j_{3}}\prod_{p=1}^{3}d_{j_{p}}w^{l}(J_{p},j_{p},\beta)\NineJ{j^{-}_{1}}{j^{-}_{2}}{j^{-}_{3}}{j^{+}_{1}}{j^{+}_{2}}{j^{+}_{3}}{j_{1}}{j_{2}}{j_{3}}^{2}
\end{equation}
Upon setting $j^{\pm}_{i} = j^{\pm}$ the reduced degenerate propagator reads:
\begin{equation}
\mathcal{K}^{j^{-}j^{+}}(i^{\pm},i'^{\pm},l,\beta) = \delta_{i^{\pm}j^{\pm}}\delta_{i'^{\pm}j^{\pm}}\sum_{j_{1}j_{2}j_{3}}
\prod_{p=1}^{3}d_{j_{p}}w^{l}(j^{-},j^{+},j_{p},\beta)
\NineJ{j^{-}}{j^{-}}{j^{-}}{j^{+}}{j^{+}}{j^{+}}{j_{1}}{j_{2}}{j_{3}}^{2}
\end{equation}
The above formulas are completely general. For the euclidean EPRL model they can be further simplified yielding the equations $A.14-A.16$ provided in Appendix, as we are going to use in the following.

\

Let us now focus on the \textit{Duflo model}. In this case the expression (\ref{I221New}) is still too complicated to be evaluated exactly, even numerically, as it stands. In order to simplify it, we use the asymptotic formula for the $9J-$symbol (\ref{NineJAsy}). Upon introducing a new coefficient $\Omega_{al}$
\begin{align}
&\Omega_{al}(j^{-},j^{+},\beta)\equiv\sum_{j=|j^{-}-j^{+}|}^{j^{-}+j^{+}}(2j+1)^{a}
w^{l}_{Duflo}(j^{-},j^{+},j,\beta) \label{OmegaNew}  
\end{align}
we can rewrite the master integral as follows:
\begin{align}
&\mathscr{I}_{Duflo}(\mathcal{G}_{221},l,\beta,\Lambda) = 
\sum_{\mathrm{All}\;j^{-}}^{\Lambda}\sum_{\mathrm{All}\;j^{+}}^{\Lambda}
\frac{d_{j^{+}_{1}}}{d_{j^{-}_{1}}d_{j^{-}_{9}}d_{j^{+}_{3}}d_{j^{+}_{4}}d_{j^{+}_{8}}}
\SixJ{j^{-}_{7}}{j^{-}_{3}}{j^{-}_{1}}{j^{-}_{9}}{j^{-}_{4}}{j^{-}_{8}}^{2}
\SixJ{j^{+}_{7}}{j^{+}_{3}}{j^{+}_{1}}{j^{+}_{9}}{j^{+}_{4}}{j^{+}_{8}}^{2} \nonumber \\
&\Omega_{0l}^{2}(j^{-}_{1},j^{+}_{1},\beta)\Omega_{0l}^{2}(j^{-}_{7},j^{+}_{7},\beta)
\Omega_{0l}(j^{-}_{3},j^{+}_{3},\beta)\Omega_{1l}(j^{-}_{3},j^{+}_{3},\beta)
\Omega_{0l}(j^{-}_{4},j^{+}_{4},\beta) \nonumber \\
&\Omega_{1l}(j^{-}_{4},j^{+}_{4},\beta)
\Omega_{0l}(j^{-}_{8},j^{+}_{8},\beta)\Omega_{1l}(j^{-}_{8},j^{+}_{8},\beta)
\Omega_{0l}(j^{-}_{9},j^{+}_{9},\beta)\Omega_{1l}(j^{-}_{9},j^{+}_{9},\beta)
\label{I221F}
\end{align}
The coefficient $\Omega_{al}$ can be easily tabulated using the analytic formula for the coefficient $w_{Duflo}$ (see eq.\,\ref{WDuflo}). The master integral (\ref{I221F}) can now be numerically evaluated\footnote{The main limitation on range of cutoff values we can test depends on the computational resources available. Here we choose $\Lambda_{\mathrm{max}} = 16$, $l=1,2$ and multiple values of $\beta$. When possible, we check for stability under extension of the range.} as a function of the cutoff for different values of the parameters $\beta$ and $l$. 

In the case of the EPRL model, instead, the general formula (\ref{I221New}) simplifies rather drastically\footnote{The $9J-$symbol in the propagator (\ref{KEffEPRL}) depends only on three spins, and one can use an equilateral formula, rather than the non-equilateral asymptotic formula (\ref{NineJAsy}) as for the Duflo model. Notice also that in the EPRL case there is no need to specify $l$, since the simplicity constraints act as a projector.}, yielding the following result
\begin{align}
&\mathscr{I}_{\mathrm{EPRL}}(\mathcal{G}_{221},l,\beta,\Lambda) = 
\sum_{\mathrm{All}\; j^{+}}^{\Lambda}
d_{j^{+}_{1}}d_{j^{+}_{3}}d_{j^{+}_{4}}d_{j^{+}_{7}}d_{j^{+}_{8}}d_{j^{+}_{9}}
d_{|\beta|j^{+}_{1}}d_{|\beta|j^{+}_{3}}d_{|\beta|j^{+}_{4}}
d_{|\beta|j^{+}_{7}}d_{|\beta|j^{+}_{8}}d_{|\beta|j^{+}_{9}} \nonumber \\
&\times\mathcal{K}^{j^{+}_{3}j^{+}_{1}j^{+}_{7}}(\beta)
\mathcal{K}^{j^{+}_{4}j^{+}_{7}j^{+}_{8}}(\beta)
\mathcal{K}^{j^{+}_{8}j^{+}_{9}j^{+}_{3}}(\beta)
\mathcal{K}^{j^{+}_{9}j^{+}_{1}j^{+}_{4}}(\beta)
\SixJ{j^{+}_{7}}{j^{+}_{3}}{j^{+}_{1}}{j^{+}_{9}}{j^{+}_{4}}{j^{+}_{8}}^{2}
\SixJ{|\beta|j^{+}_{7}}{|\beta|j^{+}_{3}}{|\beta|j^{+}_{1}}{|\beta|j^{+}_{9}}{|\beta|j^{+}_{4}}{|\beta|j^{+}_{8}}^{2}
\label{I221EPRL}
\end{align}
where the (degenerate) propagator $\mathcal{K}_{\mathrm{EPRL}}$ is given by the Eq.\,(\ref{KEffEPRL}).

\noindent A sample of the results of the numerical evaluation of the relevant master integrals\footnote{They will be discussed in detail, alongside the results for other values of the various parameters, in a follow-up publication \cite{Finocchiaro1}.} is in Figs.\,(\ref{T221Full}).
\begin{table}[t]
\centering
\begin{tabular}{cc}
\includegraphics[scale=0.65]{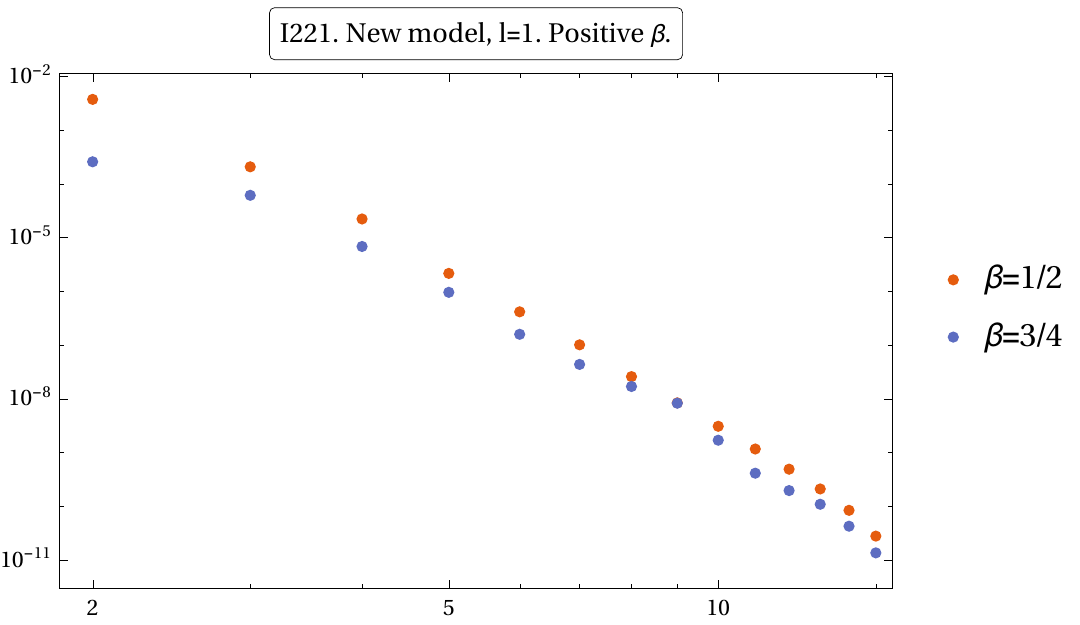} & \hspace{20pt}
\includegraphics[scale=0.65]{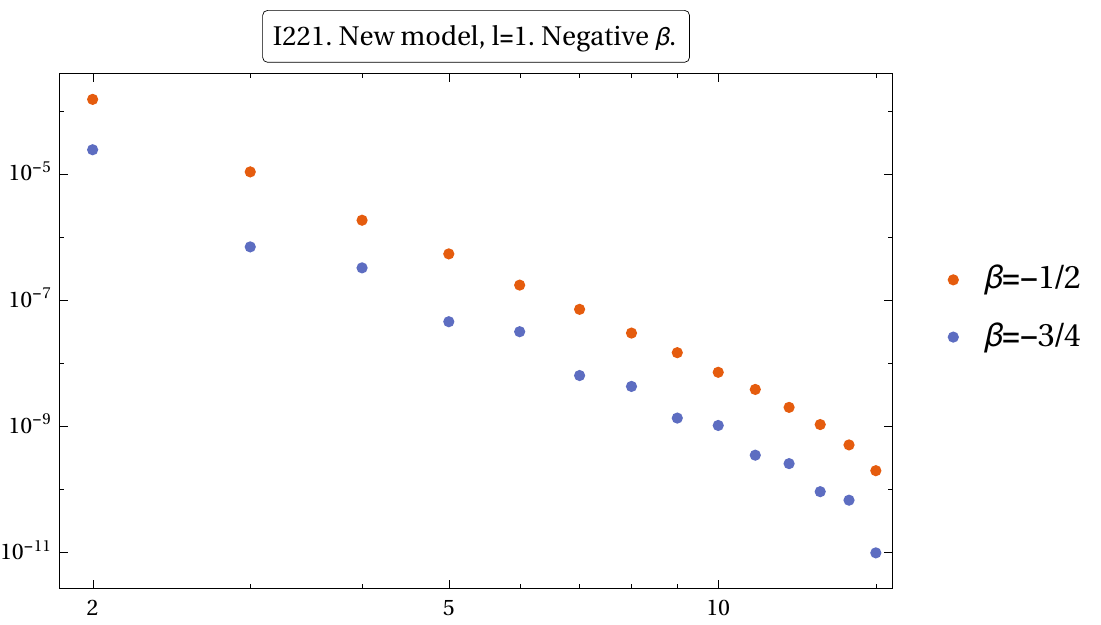} \\
\includegraphics[scale=0.65]{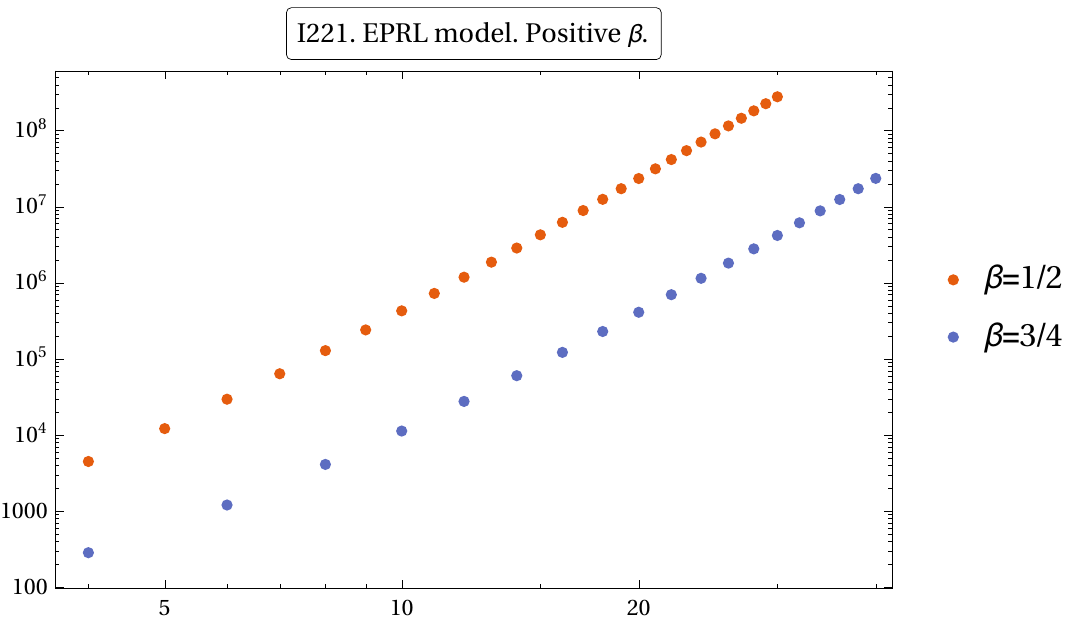} & \hspace{20pt}
\includegraphics[scale=0.65]{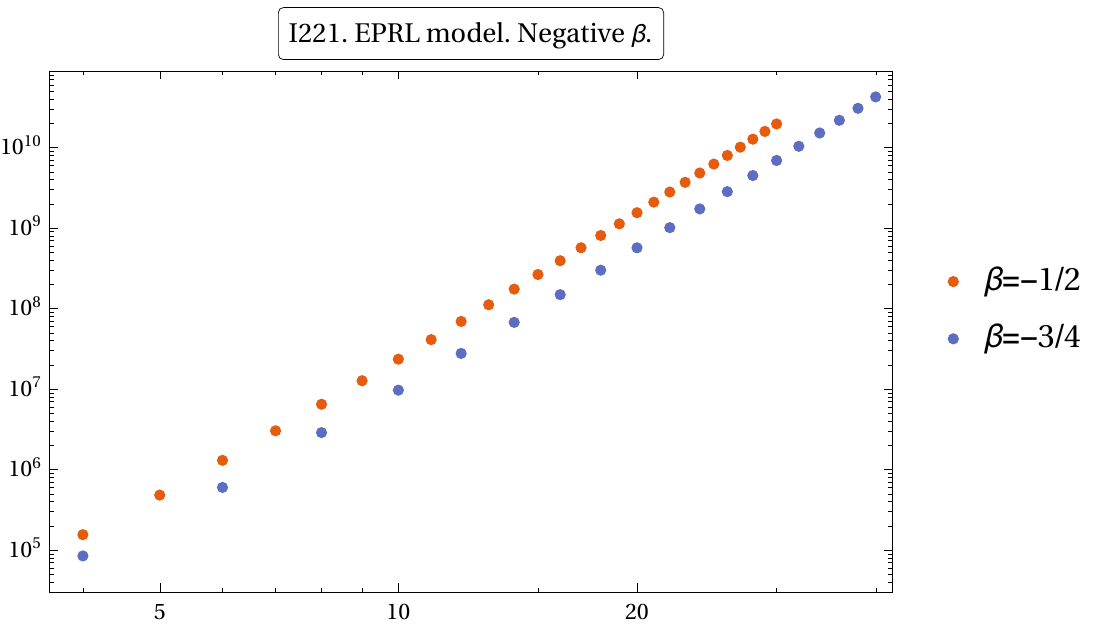}
\end{tabular}
\caption{Top: numerical evaluation of the expression $\mathscr{I}_{221}(\Lambda_{\mathrm{max}}) - \mathscr{I}_{221}(\Lambda)$ as a function of the cutoff $\Lambda\in\,[2,16]$ in a Log-Log scale for different values of $\beta$. Here $\mathscr{I}_{221}$ is given by the master integral (\ref{I221F}). Bottom: numerical evaluation of the EPRL model's master integral (\ref{I221EPRL}) as a function of the cutoff $\Lambda$ and $\beta$.
The cutoff ranges are $\Lambda\in\,[4,30]$ for $\beta=\pm\frac{1}{2}$ and $\Lambda\in\,[4,40]$ (only even values) for $\beta=\pm\frac{3}{4}$.}
\label{T221Full}
\end{table}

\noindent The degree of divergence $\omega(\mathcal{G}_{221})$ is given by the angular coefficient of the linear best fit of the data plotted in logarithmic scale. 
\noindent The mean value\footnote{To derive an estimate for $\omega $ we took the statistical average of the values obtained by fitting the data points in the cutoff ranges $\Lambda\in[\tilde{\Lambda},\Lambda_{\mathrm{max}}]$ with $\tilde{\Lambda} = 7$. The lower threshold for $\tilde{\Lambda}$ corresponds to the point from where the value of $\omega$ appears to be stable within a $10\%$ error margin (i.e. the digits to the left of the decimal point are steady).} of the scaling exponent $\bar{\omega}$ and its standard deviation for the studied cases are summarized in the table (\ref{I221Scaling}). 

\begin{table}[t]
\centering
\begin{tabular}[t]{lcccc}
\hline \\
$\mathscr{I}_{221}$ & \multicolumn{4}{c}{Numerical estimates of the degree of divergence\,\, $\bar{\omega} \pm \sigma$} \\
\hline \\
$\beta$ & $\frac{1}{2}$ & $\frac{3}{4}$ & $-\frac{1}{2}$ & $-\frac{3}{4}$ \\\vspace*{-7pt}
\\
\hline \\
Duflo model, $l=1$.\hspace{10pt} &-11.50 $\pm$ 0.85 &-11.40 $\pm$ 0.51 &-8.83 $\pm$ 1.06 &-10.40 $\pm$ 1.94 \\
Duflo model, $l=2$. &-10.77 $\pm$ 0.93 &-10.99 $\pm$ 0.82 &-10.23 $\pm$ 0.97 &-10.17 $\pm$ 0.96 \\
Duflo model, $l=1$. Asym.	& -2.40	& -1.88	& -2.28	 & -2.20 \\
Duflo model, $l=2$. Asym.	& -13.24 & -13.4 & -13.56 & -13.72 \\
EPRL model. &6.10 $\pm$ 0.04 &5.95 $\pm$ 0.06 &6.26 $\pm$ 0.026 &6.29 $\pm$ 0.035 \\
EPRL model. Asym	& 6	& 6	& 6	& 6 \\
\hline
\end{tabular}
\caption{Summary of the estimated scaling exponent $\bar{\omega}$ for leading order melonic graph $\mathcal{G}_{221}$. The EPRL values we found are in excellent agreement with other analytical results already available in the literature \cite{Geloun:2010vj}.}
\label{I221Scaling}
\end{table}\noindent

\

\noindent In the case of the Duflo model, some further subtlties arise in the evaluation, due to the more involved nature of the simplicity or geometricity coefficients. These subtleties require additional care in the numerical evaluation of scaling exponents, which are worth emphasizing here, since they are of more general validity in this class of spin foam amplitudes. The master integral formula (\ref{I221F}) relies on the use of the asymptotic formula for the $9J$-symbol (\ref{NineJAsy}). This might not be very accurate for relatively small spins, like the ones we can concretely explore in our numerical evaluations. Furthermore, since the (\ref{NineJAsy}) has a stronger suppression rate than other approximate formulas for the $9J$-symbol used in scaling analyses, e.g. the equilateral one $\{9j\}\,\approx\,j^{-\frac{8}{3}}$ (in which also all the $j$s are identified, which we cannot do in the Duflo model), the full expression (\ref{I221F}) is expected to provide \textit{only} a lower bound for $\bar{\omega}$. To test this expectation and also to cross-check the known EPRL results, obtained using the equilateral scaling, we repeat our analysis using in both cases a different asymptotic formula. \medskip\\
\noindent In order to derive an asymptotic formula for the melonic master integral $\mathscr{I}_{221}$, we localize its expression (\ref{I221New}) around a background configuration $(j^{-},j^{+})$ by setting $j^{-}_{i}=j^{-}$ and $j^{+}_{i} = j^{+}$. For the EPRL model this also implies an homogeneous identification of all the spins $j_{i}$ due to the peculiar form of the EPRL's simplicity coefficients. This procedure has been applied and tested for a number of different simplicial models, including a different version of the EPRL model \cite{Perini:2008pd}, and it seems to be reliable \cite{Riello:2013bzw, Chen:2016aag, Dona:2018nev, Dona:2018pxq, Dona:2019dkf}. 

\ 

\noindent After appropriate simplifications, the general formula (\ref{I221New}) becomes:
\begin{align}
\mathscr{I}_{\mathrm{asy}}(\mathcal{G}_{221},l,\beta,\Lambda)\,\simeq\,\Lambda^{5\mu}\sum^{\Lambda}_{j^{-}=0}\sum^{\Lambda}_{j^{+}=0}d^{6}_{j^{-}}d^{6}_{j^{+}}
\left[\mathcal{K}^{j^{-}j^{+}}(l,\beta)\right]^{4}
\SixJ{j^{-}}{j^{-}}{j^{-}}{j^{-}}{j^{-}}{j^{-}}^{2}
\SixJ{j^{-}}{j^{+}}{j^{+}}{j^{+}}{j^{+}}{j^{+}}^{2}
\label{I221Asy}
\end{align}
where $\mu=1,2$, for the EPRL and Duflo models. The reduced propagator $\mathcal{K}^{j^{-}j^{+}}$ is given by the Eq.\,(\ref{KEffUni}). To determine the amplitude's degree of divergence we combine the scaling of the various factors. The scaling of the equilateral $6J$-symbol is given by the Regge formula $\{6j\}\approx j^{-\frac{3}{2}}$. According to our analysis, in the large-$j$ regime the propagator (\ref{KEffUni}) can be very well approximated by the following expressions:
\begin{align}
\mathcal{K}^{j^{-},j^{+}}_{Duflo}(l,\beta)\,\simeq\,\frac{\delta_{j^{-}\,|\beta|j^{+}}}{(2j^{+}+1)^{\alpha}} \qquad \mathcal{K}^{j^{-}j^{+}}_{\mathrm{EPRL}}(\beta)\,\simeq\,\frac{\delta_{j^{-}\,|\beta|j^{+}}}{(2j^{+}+1)^{\frac{3}{2}}} \qquad \alpha = \alpha(l,\beta)
\label{KEffAsy}
\end{align}
The EPRL formula can be analytically derived, as shown in Appendix\,\ref{AppA} (see \ref{KEffEPRLAsy1},\,\ref{KEffEPRLAsy2}). The corresponding formula for the Duflo model follows from a direct numerical evaluation of the propagator. It is worth noticing that also the Duflo propagator, containing the much more involved Duflo geometricity coefficients (\ref{WDuflo}), peak on the same configurations of $Spin(4)$ representations $(j^+,j^-)$. This is due to the asymptotic behaviour of such coeffcients, which contains also (but not only) the configurations corresponding to the EPRL configurations among the dominant ones \cite{CFONumerical}. 

\noindent The computed values of the scaling exponent $\alpha$ can be found in the table (\ref{AlphaVal}).
\begin{table}[h]
\centering
\begin{tabular}[h]{c}
\includegraphics[scale=0.68]{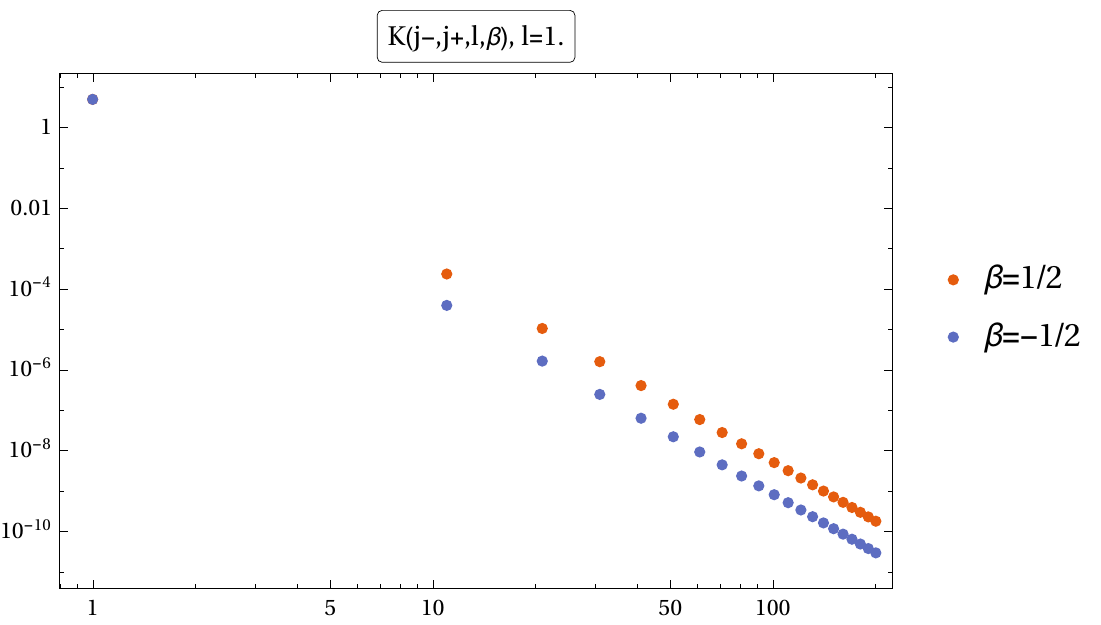}
\end{tabular}
\hspace{20pt}
\begin{tabular}[h]{ccccc}
\hline \\
\multicolumn{5}{c}{Numerical values of $\alpha$} \\
\hline \\
l	& $\frac{1}{2}$	& $\frac{3}{4}$  & $-\frac{1}{2}$ & -$\frac{3}{4}$ \\\vspace*{-8pt}
\\
\hline \\
$l=1$	& 4.85	 & 4.72 & 4.82 & 4.80 \\
$l=2$	& 7.56	 & 7.60 & 7.64 & 7.68 \\
\hline
\end{tabular}
\caption{Left panel: numerical evaluation of the propagator $\mathcal{K}_{Duflo}$ with $l=1$ and $j^{-}=|\beta|j^{+}$ in a logarithmic scale. Right panel: computed estimates of the propagator's scaling exponent $\alpha$ for various values of $\beta$ and $l$.} 
\label{AlphaVal}
\end{table}\\
The values of $\omega$, obtained by substituting the identities (\ref{KEffAsy}) into the master integral formula (\ref{I221Asy}), are listed in table (\ref{I221Scaling}). 

\

\noindent To summarize: the leading order (melonic) correction to the self-energy (\ref{I221New}) appears to be convergent for the Duflo model and divergent for the EPRL one. The degree of divergence we computed for the EPRL model is in excellent agreement with known analytical results in the literature \cite{Geloun:2010vj, Perini:2008pd}. Concerning the \textit{Duflo model} for the case $l=1$ the data clearly indicates that the use of the non-equilateral formula (\ref{NineJAsy}) in (\ref{I221F}), as appropriate for this model, strongly suppresses the amplitude scaling leading to a more convergent result. This might also be true for the case $l=2$ although we cannot state it with full confidence at the moment based on the small cutoff range we tested (recall that $\Lambda_{\mathrm{max}}=16$ in the full formula and $\Lambda_{\mathrm{max}}=100$ in the asymptotic formula). The limited range of $\Lambda$ values we explored might also explain why the difference 
between the values of $\omega$ for $l=1,2$ computed from (\ref{I221F}) (first and second row of Tab.\,\ref{I221Scaling}) is smaller than the difference between the corresponding values obtained from the asymptotic formula (third and fourth row of Tab.\,\ref{I221Scaling}).

\subsubsection[The four-point function.]{The $4$-point function.}
\begin{table}[t]
\centering
\begin{tabular}{cc}
\includegraphics[scale=0.65]{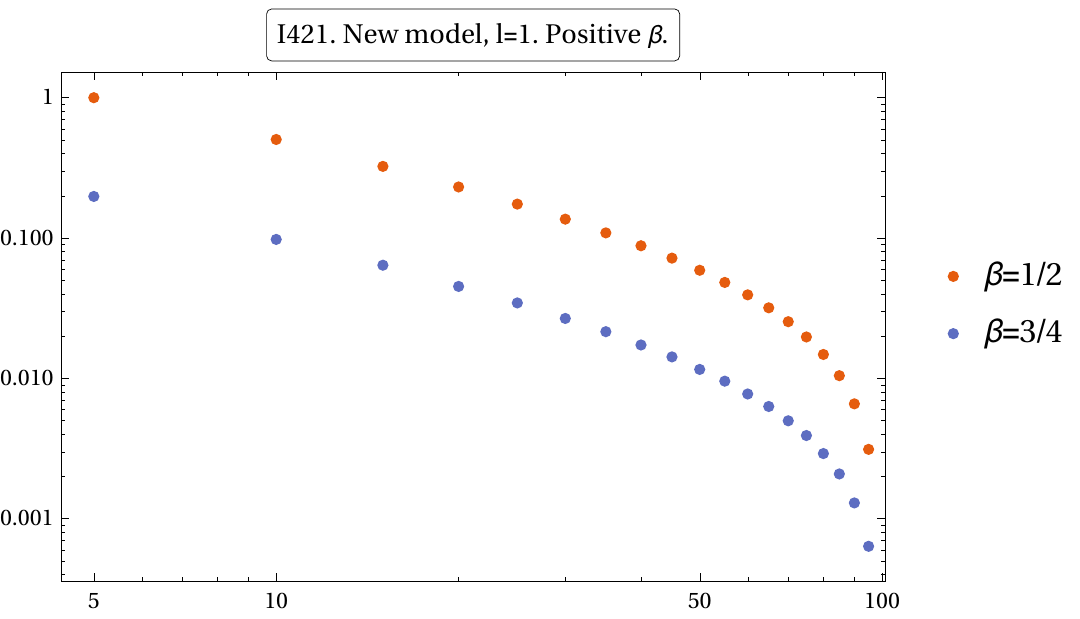} & \hspace{20pt}
\includegraphics[scale=0.65]{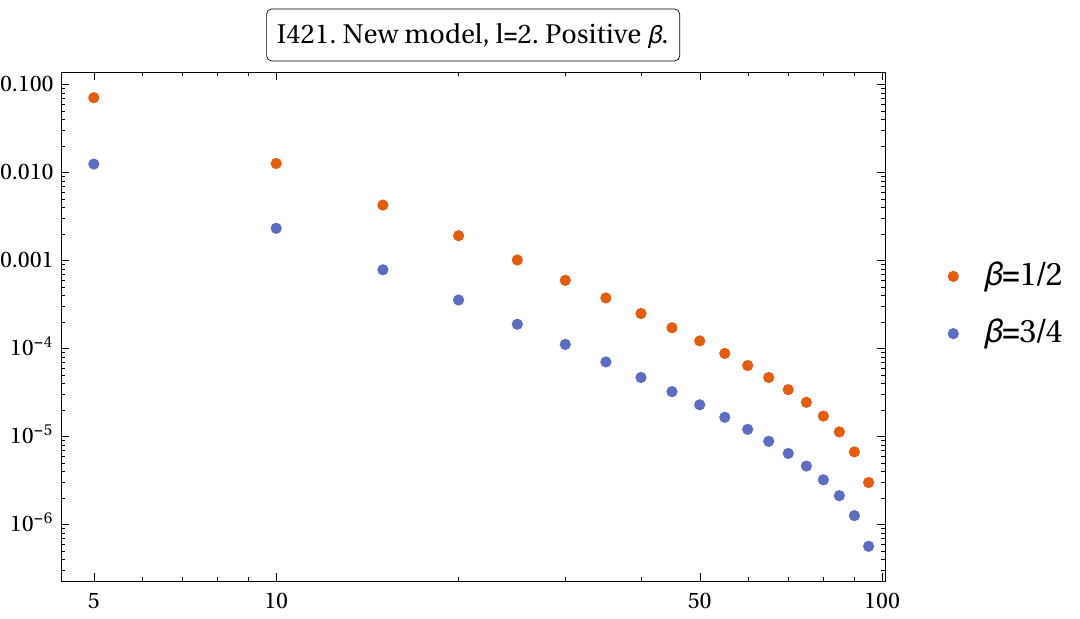} \\
\end{tabular}
\caption{Numerical evaluation of the expression $\mathscr{I}_{221}(\Lambda_{\mathrm{max}}) - \mathscr{I}_{221}(\Lambda)$ as a function of the cutoff $\Lambda\in\,[5,100]$ in a Log-Log scale for different values of $l$ and $\beta$. Here $\mathscr{I}_{421}$ is given by the Master integral formula (\ref{I421FO}).}
\label{T421}
\end{table}
We now focus on the leading (melonic) correction to the $4$-point function. The corresponding 
GFT Feynman diagram $\mathcal{G}_{421}$ is depicted in table (\ref{TabG2G4LONew}).
In order to derive the master's integral expression we follow the same strategy used in the previous section. After setting to zero the spins labelling the external faces and performing the appropriate simplification we find:
\begin{align}
\mathscr{I}(\mathcal{G}_{421},l,\beta,\Lambda) = \sum_{j_{1}^{-},\,j_{1}^{+}}^{\Lambda}
\sum_{j_{31}j_{41}j_{51}}\frac{d_{j_{31}}d_{j_{41}}d_{j_{51}}}{d_{j^{-}_{1}}d_{j^{+}_{1}}}w^{l}(j^{-}_{1},j^{+}_{1},j_{31},\beta)
w^{l}(j^{-}_{1},j^{+}_{1},j_{41},\beta)w^{l}(j^{-}_{1},j^{+}_{1},j_{51},\beta)
\label{I421}
\end{align}

\noindent Once more, the above formula is completely general and valid for any simplicial GFT (spin foam) model for constrained BF theory. 

For the EPRL and Duflo models it specializes to:
\begin{align}
&\mathscr{I}_{Duflo}(\mathcal{G}_{421},l,\beta,\Lambda) = \sum_{j_{1}^{-},\,j_{1}^{+}}^{\Lambda}
\frac{1}{d_{j^{-}_{1}}d_{j^{+}_{1}}}\left[\Omega_{1l}(j^{-}_{1},j^{+}_{1},\beta)\right]^{3} \label{I421FO} \\
&\mathscr{I}_{\mathrm{EPRL}}(\mathcal{G}_{421},\beta,\Lambda) = \sum_{j_{1}^{-}=0}^{\Lambda}\frac{(2(1-\beta)j^{+}_{1} + 1)^{3}}{(2j^{+}_{1} + 1)(2|\beta|j^{+}_{1} + 1)} \label{I421EPRL}
\end{align}
where in the first expressions we used the same notations of (\ref{I221F}). 

\noindent The degree of divergence of the master integral (\ref{I421FO}) can be computed again by fitting the data. The resulting values of $\omega$ are reported in Tab.\,(\ref{I421Scaling}).
\begin{table}[h]
\centering
\begin{tabular}[h]{lcc}
\hline \\
$\mathscr{I}_{421}$ & \multicolumn{2}{c}{Numerical estimates of the degree of divergence\,\, $\bar{\omega} \pm \sigma$} \\
\hline \\
$\beta$ & $\pm\frac{1}{2}$ & $\pm\frac{3}{4}$ \\\vspace*{-8pt}
\\
\hline \\
Duflo model, $l=1$.\hspace{10pt} & -2.36 $\pm$ 0.24 & -2.35 $\pm$ 0.24 \\
Duflo model, $l=2$. & -3.67 $\pm$ 0.20 & -3.67 $\pm$ 0.20 \\
\hline
\end{tabular}
\caption{Numerical values of the divergence's degree for the $4$-point amplitude $\mathscr{I}_{Duflo}$.}
\label{I421Scaling}
\end{table}\\\noindent
The scaling of the EPRL $4$-point amplitude can be directly read off from the corresponding formula (\ref{I421EPRL}).
\begin{equation}
\omega(\mathcal{G}_{421}) = 2 \qquad \beta\neq\,0,\,\pm 1
\end{equation}

\

\begin{figure}[t]
\centering
\includegraphics[scale=0.3,angle=-1.5,origin=c]{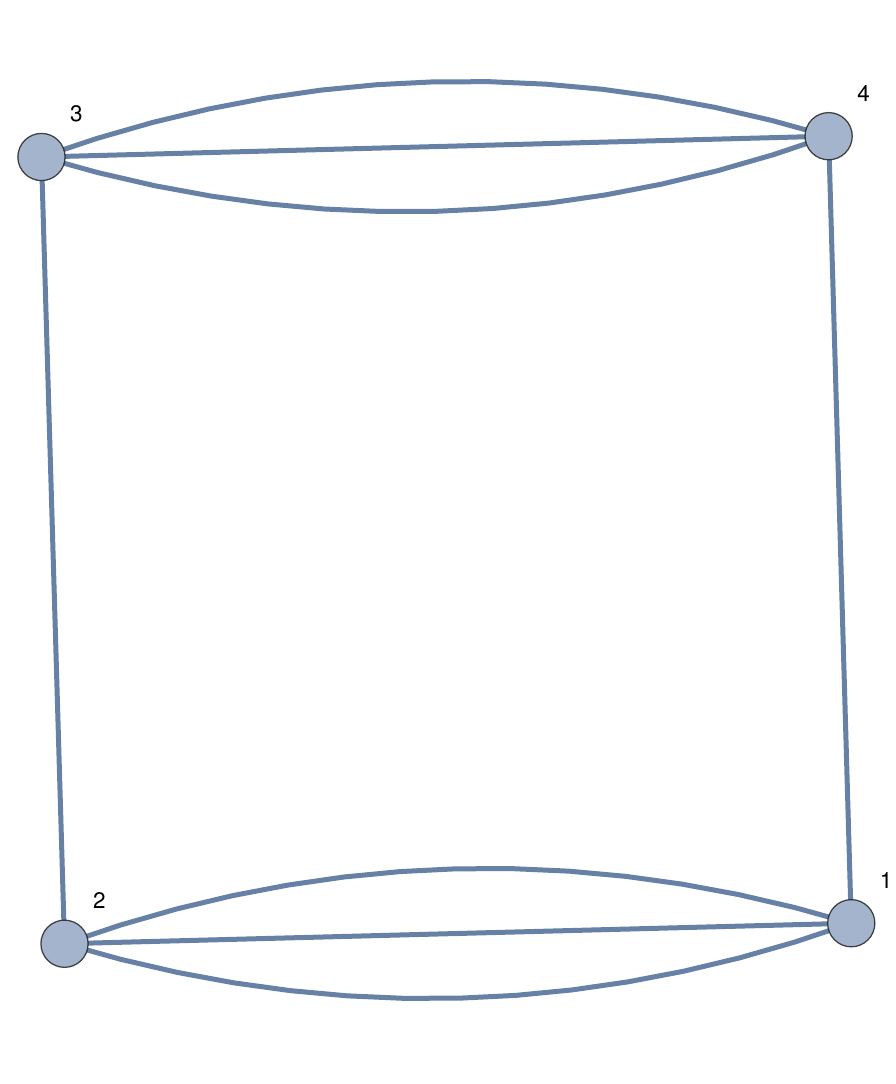}
\caption{The quartic tensorial bubble interaction $B_{41}$.}
\label{B41Vertex}
\end{figure}

To summarize: the leading order radiative correction to the $4$-point function converges for the Duflo model while it diverges quadratically in the EPRL model. 
Neglecting possible ambiguities in the definition of both models (which, as we emphasized earlier, could affect the face amplitudes and thus the precise scaling behaviour) it would then seem that the Duflo model does not require renormalization, at least at this order, while the EPRL model does. But of course higher orders are needed to establish the renormalizability of both models, thus it is hard to draw too many conclusions from this result.
\noindent More than the divergence degree in itself, it is important to notice that, since the $\mathcal{G}_{421}$ is melonic (and thus tracial), the corresponding counterterm that is required to absorb the divergence, when present, is proportional to a tensor invariant quartic interaction term (more precisely to the bubble $B_{41}$ vertex in Fig.\,\ref{B41Vertex}). 
Such counterterm is incompatible with a pure simplicial theory space (e.g. a strictly simplicial EPRL model would be non-renormalizable), and this signals the need to extend the theory space of geometric GFT models beyond the simplicial ansatz to include tensorial bubble interactions.
\subsubsection[Next-to-leading order corrections to the N-point functions.]{Next-to-leading order corrections to the $N$-point functions.}
We now show how to generalize the master integral formulas for the melonic $2$-point function $\mathcal{G}_{221}$ to an important class of higher order GFT Feynman diagrams.

\noindent The master integral expressions (\ref{I221New},\,\ref{I221Asy}) rely on the property that each internal link of the diagram is shared exactly by one external face and three internal ones. Hence, after setting to zero the spins labelling the external faces we are left with:
i) a pair of $6J$-symbols for each vertex (coming from a pair of degenerate $15J$-symbols);
ii) propagator of the form (\ref{KEff}) for each internal edge. A pair of face weights $d_{j^{-}_{i}}d_{j^{+}_{i}}$ for each internal face.

\begin{figure}[t]
\centering
\begin{tabular}[t]{cc}
\hline \\
\multicolumn{2}{c}{NLO corrections to the $3$-point and $4$-point functions.}\\
\hline \\
\includegraphics[scale=0.55]{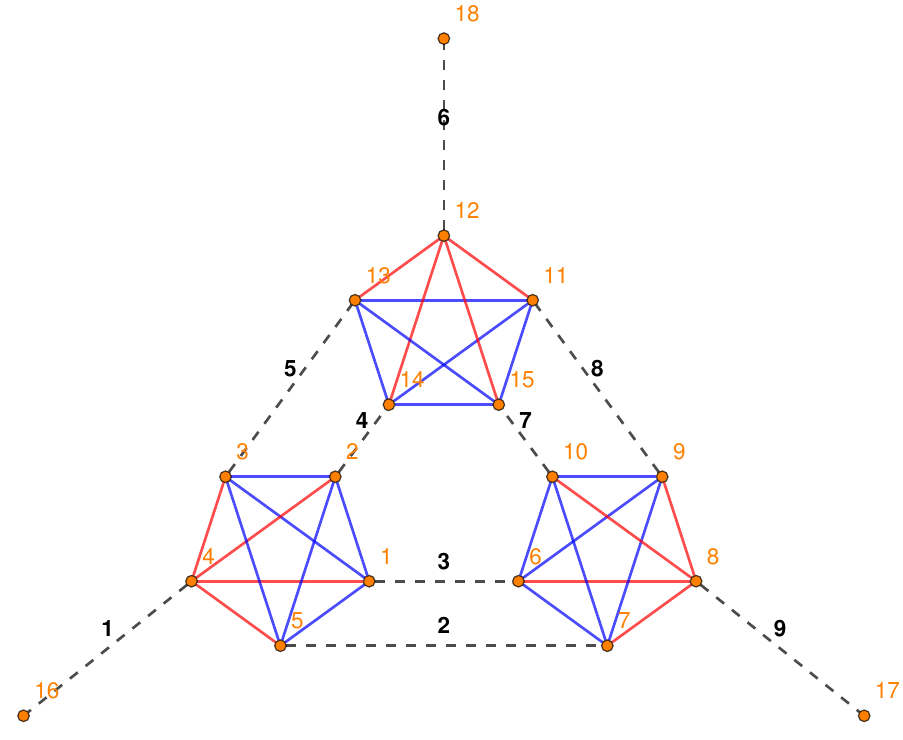} & \includegraphics[scale=0.45]{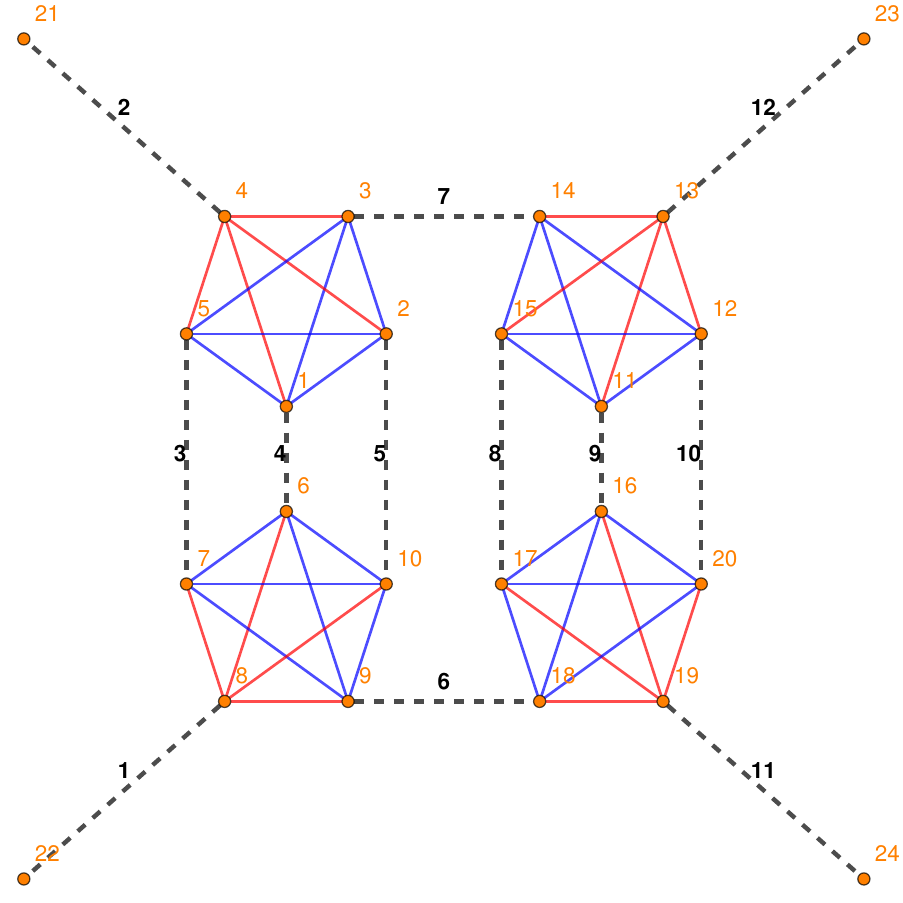} \\ \includegraphics[scale=0.45]{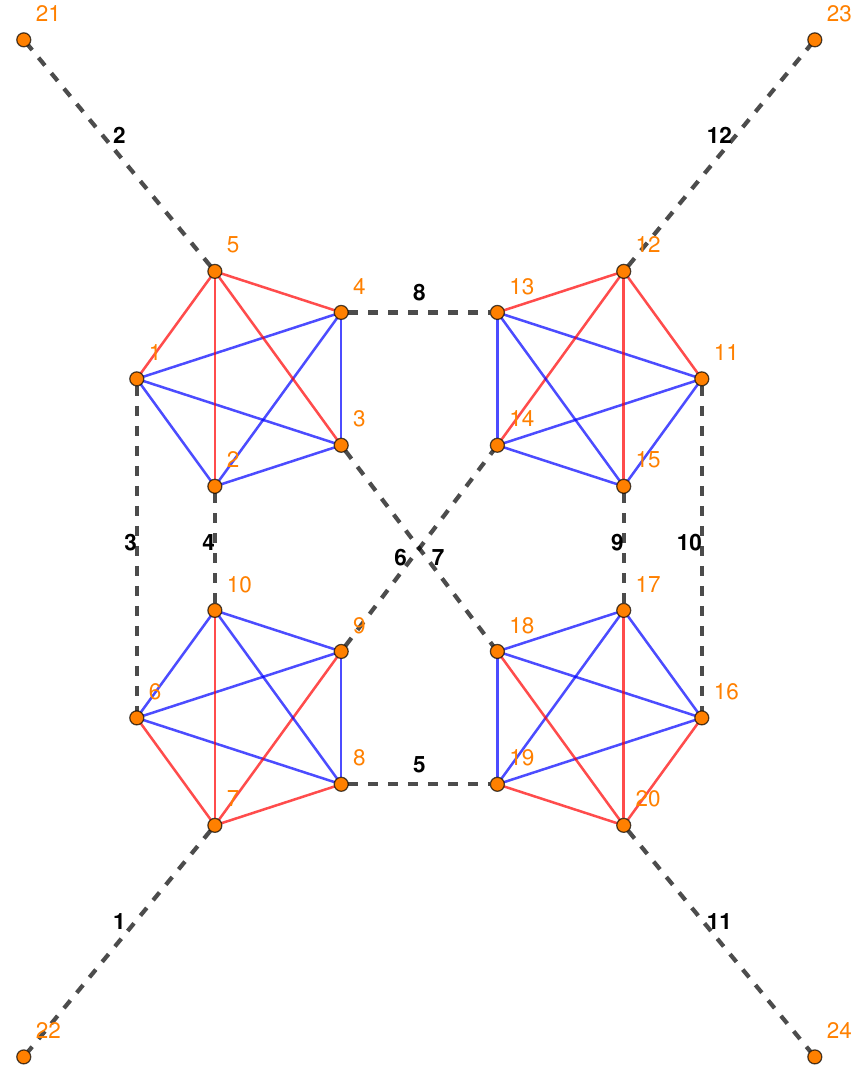} & \includegraphics[scale=0.45]{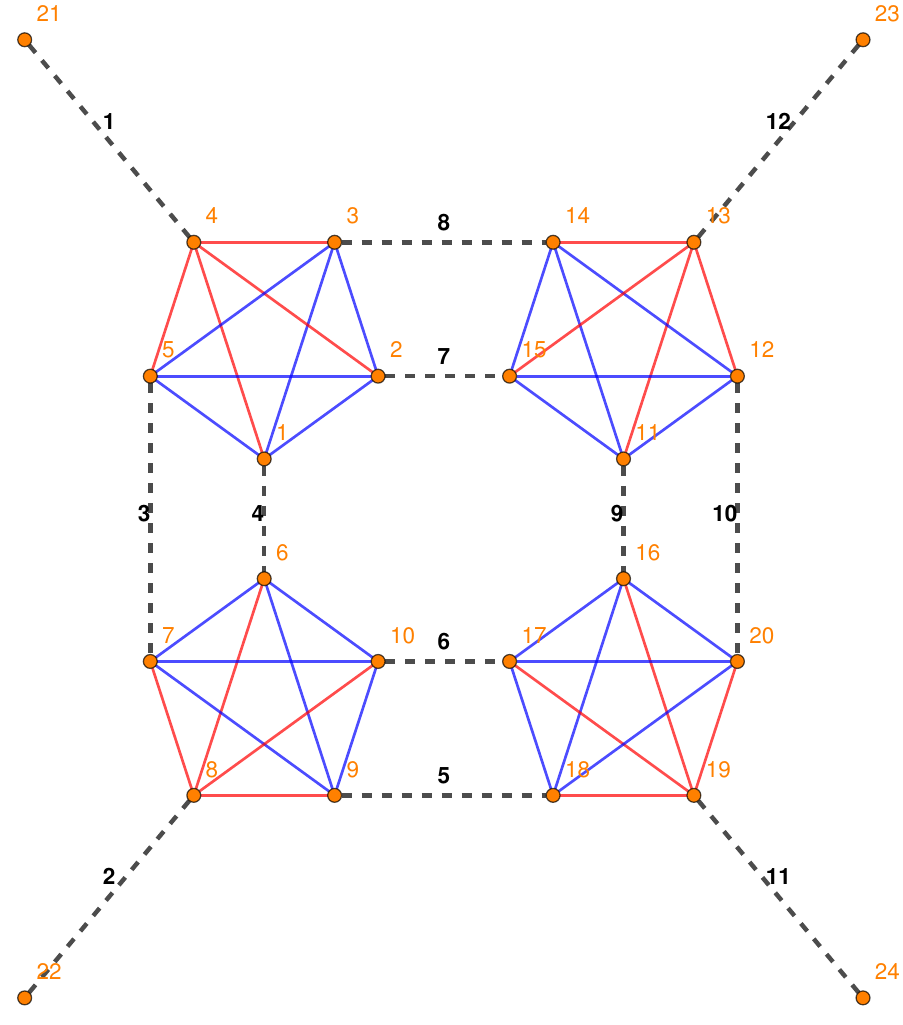} \\
\hline
\end{tabular}
\caption{Example of GFT diagrams of the type $\mathcal{G}_{N_{v}N_{v}n}$ contributing to the $3-$ and $4-$point functions at the NLO.}
\label{TabG3G4NLO}
\end{figure}

\noindent The above combinatorial property is true for any tadpole-free GFT diagram with one and only one external link for each simplicial vertex (here denoted as $\mathcal{G}_{N_{v}N_{v}n}$). Some examples of these diagrams are shown in Fig.\,(\ref{TabG3G4NLO}). Therefore the expression (\ref{I221Asy}) can be generalized as follows:
\begin{align}
&\mathscr{I}(\mathcal{G}_{N_{v}N_{v}n},l,\beta,\Lambda) = \sum_{J_{f}\in\mathcal{F}_{\mathrm{cl}}}^{\Lambda}\prod_{f\in\mathcal{F}_{\mathrm{cl}}}
d_{J_{f}}\prod_{e\in\mathcal{E}_{\mathrm{int}}}\mathcal{K}^{J_{f_{i}}J_{f_{j}}J_{f_{k}}}(l,\beta)\prod_{v\in\mathcal{V}}\{6J_{vf}\} \label{IG} \\
&\mathscr{I}_{\mathrm{asy}}(\mathcal{G}_{N_{v}N_{v}n},l,\beta,\Lambda) = \Lambda^{(N_{f}-1)\mu}\sum_{j^{-},\,j^{+}}^{\Lambda}d^{N_{f}}_{j^{-}}d^{N_{f}}_{j^{+}}
\left[\mathcal{K}^{j^{-}j^{+}}(l,\beta)\right]^{N_{e}}
\SixJ{j^{-}}{j^{-}}{j^{-}}{j^{-}}{j^{-}}{j^{-}}^{N_{v}}
\SixJ{j^{-}}{j^{+}}{j^{+}}{j^{+}}{j^{+}}{j^{+}}^{N_{v}} \label{IGAsy}
\end{align}
with $\mu=1,2$ respectively for the EPRL and the Duflo model. 

\noindent Upon using the identities (\ref{KEffAsy}), the asymptotic master integral (\ref{IGAsy}) takes the following form:
\begin{align}
&\mathscr{I}_{\mathrm{asy}}(\mathcal{G}_{N_{v}N_{v}n},l,\beta,\Lambda)\,\simeq\, \Lambda^{(N_{f}-1)\mu}\sum_{j^{+}}^{\Lambda}\left(j^{+}\right)^{2N_{f}-\alpha N_{e} - 3N_{v}} = \Lambda^{(\mu+2)N_{f}-3N_{v}-\alpha N_{e} -\mu +1} \\
&\omega(\mathcal{G}_{N_{v}N_{v}n}) = (\mu+2)N_{f}-3N_{v}-\alpha N_{e} -\mu +1
\end{align}
For the Duflo model the value of $\alpha$ must be computed on case by case basis (see Tab.\,\ref{AlphaVal}), while for the EPRL model we have $\alpha=\frac{3}{2}$. 

\noindent The degree of divergence $\omega$ can also be written in terms of the number of vertices $N_{v}$
\begin{equation}
\omega(\mathcal{G}_{N_{v}N_{v}n}) = (\mu+2)N_{f} - (2\alpha + 3)N_{v} -\mu + 1 
\qquad N_{e} = \frac{5N_{v} - N_{\mathrm{ext}}}{2} = 2N_{v} \label{PCAsy}
\end{equation}
For the diagrams in Fig.\,(\ref{TabG3G4NLO}) the above formula give us the following results, reported in the table (\ref{NLOScaling}): the Duflo model amplitudes for all four diagrams are finite; for the EPRL model\footnote{We point out again that we are studying both models under specific choices fixing the various ambiguities that enter the construction of the spin foam amplitudes. These ambiguities affect, in general, the scaling results.} the first and last diagrams might be logarithmically divergent and therefore require a deeper analysis \cite{Finocchiaro1}. The diagram $\mathcal{G}_{442}$ converges, while the melonic diagram $\mathcal{G}_{441}$ diverges as a cubic power of the cutoff. The corresponding counterterm is proportional to the quartic bubble vertex $B_{41}$, suggesting again the need for a suitable extension of the theory space to incorporate the relevant tensorial interactions. 

Once more, this is in fact the crucial lesson we draw from this analysis of divergences, more important, we think, than the precise scaling of the amplitudes, for the reasons already explained.
\begin{table}[t]
\centering
\begin{tabular}[t]{lcccc}
\hline \\
\multicolumn{5}{c}{Degree of divergence. NLO graphs.} \\
\hline \\
$\mathcal{G}$ & $\mathcal{G}_{331}$ & $\mathcal{G}_{441}$ & $\mathcal{G}_{442}$ & $\mathcal{G}_{443}$ \\\vspace*{-8pt}
\\
\hline \\
Duflo model, $l=1$, $\beta=\frac{1}{2}$.\hspace{10pt} & -15.1  & -15.8  & -31.8   & -19.8 \\
EPRL model & 0  & 3  & -9  & 0 \\
\hline
\end{tabular}
\caption{Degree of divergence for the diagrams in fig.\,\ref{TabG3G4NLO} computed from the power counting formula (\ref{PCAsy}).}
\label{NLOScaling}
\end{table}
\subsubsection[Beyond perturbation theory: the necklace diagrams]{Beyond perturbation theory: the necklace diagrams}
In the previous two subsections we discussed the all the leading order and a subclass of next-to-leading order radiative corrections to the $N$-point functions with $N\leq 6$. Now we show how to use some of the results we found to estimate the scaling of the so-called necklace diagrams to all orders in perturbation theory. 

\noindent A (connected) GFT Feynman diagram belongs to the necklace class if (and only if) it consists of an open chain of vertices where each vertex (except the first and last ones) is connected only to its two closest neighbours. Here we restrict to the set of $k$-necklace diagrams with $4\leq k\leq 6$ where $k$ denotes the number of external links, since all other higher order diagrams are either 1-particle reducible or manifestly convergent. Three examples of necklace diagrams are shown in Tab.\,(\ref{TabNecklace}). 
\begin{table}[t]
\centering
\begin{tabular}[t]{c}
\hline \\
GFT necklace graphs \\
\hline \\
\includegraphics[scale=1.0]{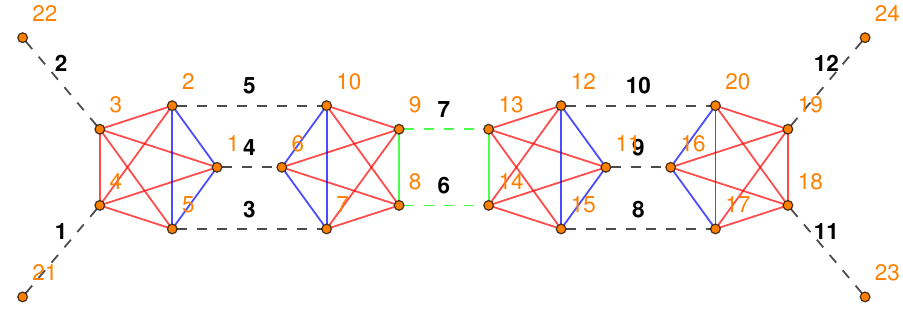} \\ 
\includegraphics[scale=1.3]{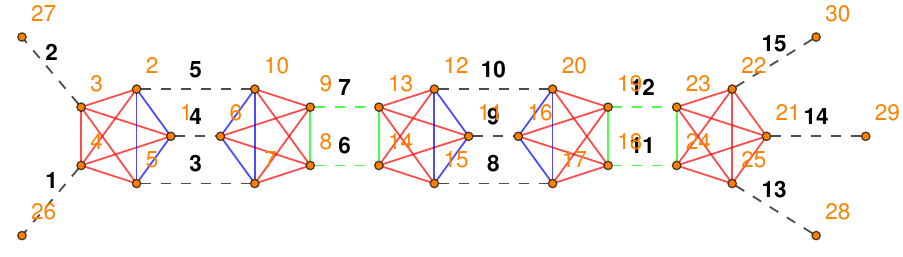} \\ 
\includegraphics[scale=1.2]{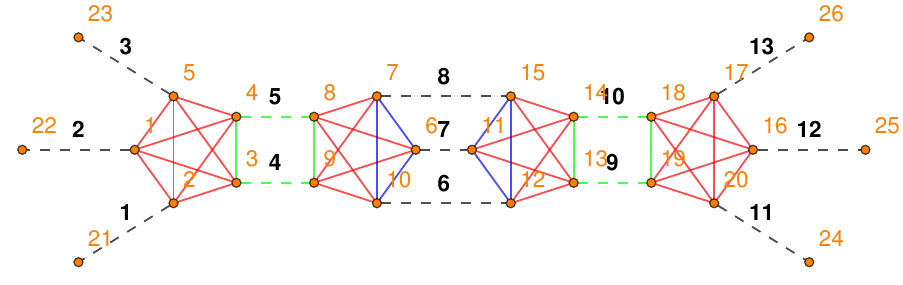} \\
\hline
\end{tabular}
\caption{Examples of necklace diagrams contributing to the $4-$, $5-$ and $6-$point functions}
\label{TabNecklace}
\end{table}
The necklace diagrams share the following remarkable property: the set of internal faces of a $k$-necklace diagram can always be decomposed into the direct sum of two subsets
\begin{align}
\mathcal{F}(\mathcal{G}_{k-\mathrm{necklace}}) = \bar{\mathcal{F}}\oplus\tilde{F} \qquad 
\bar{\mathcal{F}} = \oplus_{i=1}^{N_{b}}\mathcal{F}_{\mathcal{B}_{421}} \qquad \tilde{\mathcal{F}}=\{f_{j}\;|\;f_{j}\; \mathrm{contractible}\; \forall j\}
\end{align}
where the (three) faces in each set $\mathcal{F}_{\mathcal{B}_{421}}$ form the bubble subgraph 
$\mathcal{B}_{421}$ of the melonic $4$-point diagram $\mathcal{G}_{421}$ while the faces in $\tilde{\mathcal{F}}$ are contractible. The integer $N_{b}$, namely the number of disjoint bubbles $\mathcal{B}_{421}$ depends on the connectivity of the necklace diagram itself (i.e. on its number of vertices and external links). Thus after setting to zero the spins labelling the external and the contractible internal faces, the amplitude of $k$-necklace diagram factorizes into $N$ identical copies of the master integral $\mathscr{I}_{421}$ given in (\ref{I421}). More in detail, we have:
\begin{equation}
\mathcal{A}(\mathcal{G}_{k-\mathrm{necklace}},l,\beta,\Lambda) = \left[\mathscr{I}(\mathcal{G}_{421},l,\beta,\Lambda)\right]^{\frac{N_{v}-\mathrm{mod}(k,2)}{\lfloor k, 2\rfloor}}
\end{equation}

\

To summarize: the master integral $\mathscr{I}_{421}$ encodes the scaling of the necklace diagrams to all orders in perturbation theory. For the EPRL model a consistent (recursive) subtraction of all divergences associated to $k$-necklace graph requires an extension of the theory space to include the appropriate tensor invariant interactions of order four and six.  
A more detailed analysis of the combinatorial structure of the require counterterms is left for future work \cite{Finocchiaro1}.

\section{The road ahead: some suggestions}
We close with a set of suggestions for research directions to be pursued, toward a complete understanding the renormalization group flow of simplicial GFT models for 4d quantum gravity.

\subsection{Scaling and more scaling, and the discrete geometry of divergent configurations} 

The first suggestion is stating the obvious: compute, compute, compute\footnote{To be clear: 1-loop and 2-loop calculations would be probably enough to extract a lot of interesting properties from the perturbative expansion of GFT models for 4d quantum gravity. They may even be sufficient, since we actually expect that most interesting physics should be looked for elsewhere, i.e. within different regimes and approximations of the fundamental quantum dynamics, and not in the perturbative (spin foam) expansion. Moreover, as we stressed, such perturbative calculations should have as main goal to identify the theory space within which the relevant GFT models should be placed, more than simply pushing the perturbative analysis for its own sake. We emphasize, however, that, in fact, a lot of computational effort is required to perform such 1-loop and 2-loop calculations because one needs to consider many and complicated GFT diagrams, and involved spin foam amplitudes, already at this order. These are the calculations we intend to encourage.}. We need to (pardon us the word pun) scale up the effort in investigating the scaling of simplicial GFT amplitudes, at the same time trying to import insights from tensorial GFT amplitudes. We need to know much more about the divergent configurations and their dependence on the combinatorics of the underlying cellular complex. Lacking better tools, hard brute-force computations of spin foam amplitudes are the inevitable duty, and these in turn can only build on a better control of the relevant buliding block the vertex amplitude or, in GFT language, the vertex kernel (which plays an important role also in non-perturbative calculations). 
Brute force alone will not lead us far, however. On the one hand, we need to develop a more refined analytic understanding of these kernels and resulting amplitudes, and to identify simplified expressions that capture the relevant scaling properties, and their behaviour under coarse-graining. On the other hand, where analytical methods do not reach, we need numerical ones to take over; numerical tools for the evaluation of GFT amplitudes are thus badly needed. Packages like have been developed for local QFT Feynmanology would be of course most welcome.  
On the analytic side, another important objective should be to characterize in detail the (simplicial) geometric meaning of the dominant, most divergent configurations. This is needed to understand  the nature of the needed counterterms, but it may also provide insights on the physical features of such GFT models, even beyond their discrete formulation.

The goal here is not to so much to be able to compute GFT Feynman diagrams to arbitrary order (e.g. in vertex or loop expansions). Even in standard QFT, for the physical questions for which the perturbative expansion is the correct approximation scheme, we need to compute (very) many diagrams, but there is often no need nor possibility to go beyond some (usually low) order of approximation and beyond a certain (usually small) number of  physical degrees of freedom in the chosen boundary states (of course, the two restrictions go together, since for highly populated boundary states, even the simplest diagrams are of high order). A clear physical picture behind this approximation scheme is as important as computational power. Moreover,  coming to the specific quantum gravity case, we would argue that the perturbative GFT expansion, and the description of the dynamics in terms of elementary processes involving few of the fundamental quanta, ie. the usual spin foam language, is not the most convenient approximation to capture the effective continuum physics of quantum gravity, e.g. concerning early cosmology or quantum black holes. 

Nor the goal of such analysis of perturbative GFT divergences is establishing that one specific GFT model is finite. Not only renormalizability is a more subtle and possibly interesting feature that finiteness, but one can imagine playing with the ambiguities entering the construction of any given model to modify its scaling behaviour and turning it into a finite one. This could be a way to fix or constrain such ambiguities, of course, but it also shows that finiteness per se probably should not be a goal, and that physical conditions fixing the same ambiguities are needed.
The main goal of actual computations of GFT amplitudes should then be to provide solid indications on the general power counting of divergences, on the way to a renormalizability proof, and, even more, to indicate the relevant counterterms to be added to the model, and thus the relevant theory space of the starting GFT model. More generally, the goal of such perturbative calculations should be to provide information and tools to be employed to go beyond the perturbative setting and dwell into non-perturbative GFT renormalization. It is only the latter that can provide us with the insights and the results we need to truly explore the formal solidity and effective continuum physics of GFT models of quantum gravity.

\subsection{GFT theory space, colors and relation between simplicial and tensorial models}

We have emphasized several times already the importance of defining the relevant theory space of simplicial GFT models, in order to set up a proper renormalization scheme (perturbative and non-perturbative). Much more work should be devoted to this issue, in particular understanding more about the symmetries of such quantum gravity models.
One question is whether the yet to be identified theory space of simplicial GFT models relates to the one of tensorial GFTs. We speculate that they do and, in fact, some hints that the two may largely coincide are known. 
First, taking seriously the tensorial nature of GFT fields implies coloring (thus distinguishing and ordering) their arguments and, as a consequence, their Feynman diagrams. As we noted, this coloring allows a precise control over the topology of the cellular complexes dual to these Feynman diagrams \cite{Gurau:2012hl} and, in turn, this greater control allowed for many results that are central in renormalization analyses of tensorial GFTs (e.g. large-N expansions) \cite{Carrozza:2013mna, Carrozza:2016vsq}. A precise control over the topology of the Feynman diagrams, i.e. the cellular complexes on which spin foam amplitudes are based, is needed also in simplicial GFTs, if one aims at identifying the nature of divergences, leading to precise power counting results. It is also needed for identifying key symmetries, as we know already in the case of topological BF models \cite{Bonzom:2012mb}, where the complete power counting also relied on the full topological information on the underlying cellular complex \cite{Bonzom:2011br}. Thus we have a strong argument for relying on colors also in simplicial GFT models of 4d quantum gravity; the form of the corresponding spin foam amplitudes would remain unchanged, but they would now be defined on full 4d cellular complexes, rather than just their 2-skeleton. 
Assuming we work on colored simplicial GFT models, we then have two preliminary results that suggest a close relation with tensorial models. One is that using colors one can identify similar symmetries in the simplicial case than one finds in the tensorial one \cite{Kegeles:2016wfg}. The other is that integrating out all fields except one in a colored simplicial GFT model (in any dimension, with trivial kinetic term) produces an equivalent tensorial GFT model for the remaining field (with the same coupling constant for all interactions) \cite{Gurau:2011tj}. 
A third general fact pointing in the same direction is that divergences in simplicial GFT models for topological BF theory, which is the starting point of the construction of simplicial 4d gravity models, are associated to bubbles in the cellular complex, which are in fact the cells associated to allowed interactions in tensorial GFTs with the same base group manifold.
These results, in our opinion, suggest that there could be a single theory space containing both (colored) simplicial models and tensorial ones, with interaction kernels in the tensorial directions yet to be identified.

\subsection{GFT models with local directions}

The third suggestion for further research is to devote attention to the renormalization of GFT models which combine the combinatorially non-local pairing structure on geometric variables, in GFT interactions, with the presence of local directions. This includes both simplicial GFT models and tensorial ones, with the distinction referring to the pairing of geometric variables. 

There are two main examples of such \lq mixed\rq models. One is the tensorial models used to describe SYK-like many-body systems \cite{Delporte:2018iyf}, whose renormalization has been in fact studied in several cases. Here the non-local, tensorial indices are usually reduced to finite sets (we have thus simple tensor models, rather than full GFTs) and the single local direction is a time variable. The standard SYK models are indeed quantum mechanical models in 0+1 dimensions, with generalizations to higher dimensions (thus, with more local directions) having been proposed. 
The other class of mixed models is the extension of (simplicial) GFT quantum gravity models to include scalar fields coupled to gravitational degrees of freedom \cite{Li:2017uao}. These extended models have been studied in particular in the context of GFT condensate cosmology \cite{Oriti:2016qtz, Wilson-Ewing:2018mrp, Gielen:2016dss, Oriti:2016acw, Gielen:2017eco}, with the additional scalar fields playing (also) the role of clock and rods that allow to define relational, diffeo-invariant observables in terms of which an effective cosmological dynamics can be extracted from the GFT hydrodynamics. 

The potential physical interest of these models, and of their renormalization analysis, is thus obvious. They present several interesting issues. The presence of both local and non-local directions may modify sensibly the renormalization flow and the structure of divergences, thus leading to different dominant diagrams and effective dynamics in both UV and IR sectors. One can also envisage setting up an altogether different renormalization group scheme, adopting a notion of scale tied to the scalar (local) directions, rather than the group manifold (or involving both), potentially producing very different results. Such focus on the flow parametrised by variables with a (tentative) physical interpretation as relational time/space variables may also allow a more direct physical interpretation of the renormalization  flow itself, e.g. in a cosmological context (even though similar cautionary remarks as for the usual renormalization scheme would apply here).

\subsection{Relation with lattice spin foam renormalization}

We have emphasized how renormalizing a GFT model is tantamount to renormalizing (and studying the continuum limit of) the corresponding discrete gravity path integral and spin foam amplitudes, from a different standpoint. But the GFT formalism is only one way to provide a complete definition of spin foam models, the other being to view them as a peculiar (because background independent) lattice theory and setting up some appropriate refinement procedure. Therefore,  it would be very important to compare results obtained in the context of GFT renormalization, especially for simplicial quantum gravity models, with the results and techniques developed for renormalizing spin foam amplitudes from a lattice gauge theory perspective \cite{Dittrich:2014ala, Dittrich:2016tys, Delcamp:2016dqo, Bahr:2017klw, Bahr:2014qza}. 

In this lattice-focused approach to spin foam renormalization, a cut-off is also imposed on representation variables, but the notion of \lq scale\rq is rather given by the combinatorial complexity of the underlying lattice, and the renormalization group flow is driven by refinement/coarse-graining steps ordered by such complexity. Refinement/coarse-graining steps affect both bulk lattices and boundary graphs, and the flow of quantum amplitudes is constrained by the requirement of their consistency under restriction to coarser boundary states. 

Despite their differences, the two renormalization schemes share several, since also GFT subtraction moves amount to lattice coarse-graining steps, and corresponding maps between associated amplitudes are also built-in in the (perturbative) QFT renormalization steps used in the GFT context. Still, a detailed work of translation between the two frameworks would be very useful. This work may require, on the GFT side, a combination of functional renormalization group techniques, since we are interested in the continuum limit of spin foam models, and perturbative expansions, given that spin foam models arise in such expansion. This comparison would be beneficial for both approaches; in particular, it would emphasize the role of combinatorial complexity of boundary states in the GFT renormalization flow.  
This work should be carried out for all models that have been studied in both settings (also in the lattice renormalization approach work has been confined mostly to highly simplified models), aiming of course at unraveling the continuum phase diagram of 4d quantum gravity from two perspectives at once.

\subsection{GFT renormalization via tensor networks}

One powerful set of techniques coming from the theory of quantum many-body systems, that have been already applied in the context of lattice-based renormalization of spin foam models, uses the language of tensor networks \cite{Dittrich:2014mxa, Cunningham:2020uco}. This language is useful both for numerical studies and for emphasizing the role of entanglement in the renormalization group flow \cite{Vidal:2007hda, Orus:2013kga}; in particular, it allows to unravel topological quantum phases of many-body systems. 

In the case of GFT models, the interest in importing techniques from tensor networks goes beyond these general facts, and stems also from the fact that GFT states themselves can be seen as generalised tensor networks \cite{Chirco:2017vhs}, and by the related fact that entanglement is responsible for the basic connectivity between GFT quanta that gives rise to extended discrete structures labeled by quantum geometric data. The many facets of the GFT formalism, moreover, would allow for a manifold application of tensor network techniques. On the one hand the basic GFT field is a tensor and its quantum states are tensor networks, as mentioned; on the other hand, it remains a QFT, calling for continuum tensor network techniques as employed, say, in standard scalar quantum field theory \cite{Hu:2018hyd}. At the same time, its Feynman amplitudes are lattice gauge theories, to which a different set of tensor network techniques can be applied \cite{Tagliacozzo:2014bta} (as developed in the context of spin foam lattice renormalization). And they remain quantum many-body systems, peculiar for their background independent nature, but still conventional enough to allow the deployment of tensor network methods taken from their natural context.    

\newpage
\appendix
\section{Useful identities.}\label{AppA}
Many of the formulae reported here can be found in (or straightforwardly derived from) the reference \cite{Varshalovich:1988ye}
\subsection[9J-symbol]{$9J$-symbol.}
In this note we only need the identities regarding two types of degenerate $9J$-symbols. 
\begin{description}[leftmargin=0pt]
\item[Type A.] One degenerate triad. This type of $9J$-symbol occurs in the evaluation of the bulk amplitudes.
\begin{align}
&\NineJ{0}{j_{2}}{j_{3}}{0}{j_{5}}{j_{6}}{0}{j_{8}}{j_{9}} = \delta_{j_{2}j_{3}}\frac{(-1)^{j_{3}+j_{6}+j_{8}}}{\sqrt{(2j_{3}+1)}}\SixJ{j_{9}}{j_{6}}{j_{3}}{j_{5}}{j_{8}}{0} 
= \delta_{j_{2}j_{3}}\delta_{j_{5}j_{6}}\delta_{j_{8}j_{9}}\frac{1}{\sqrt{(2j_{3}+1)(2j_{6}+1)(2j_{9}+1)}} \label{NineJA}
\end{align}
\item[Type B.] One row is the sum or the difference of the other two. The \textit{type B} NineJ symbol coincides with the NineJ symbol with three degenerate triads in the equation $14$, paragraph $10.8.4$ of \cite{Varshalovich:1988ye}. The type B $9J$-symbols occur in the full and asymptotic formulas of the EPRL $2$-point master integral $\mathscr{I}_{221}$. In detail, we have:
\begin{align}
&\NineJ{bj^{+}_{1}}{bj^{+}_{2}}{bj^{+}_{3}}{j^{+}_{1}}{j^{+}_{2}}{j^{+}_{3}}{aj^{+}_{1}}{aj^{+}_{2}}{aj^{+}_{3}}\;\underset{\beta < 0}{\approx}\;\left[\frac{4a}{b\pi}\right]^{\frac{1}{4}}\Delta(j^{+}_{1},j^{+}_{2},j^{+}_{3})
\left[\frac{1 + a(j^{+}_{1} + j^{+}_{2} + j^{+}_{3})}{(1 + 2aj^{+}_{1})(1 + 2aj^{+}_{2})(1 + 2aj^{+}_{3})(1 + j^{+}_{1} + j^{+}_{2} +j^{+}_{3})}\right]^{\frac{1}{2}} \label{NineJBM}\\
&\NineJ{bj^{+}_{1}}{bj^{+}_{2}}{bj^{+}_{3}}{\delta j^{+}_{1}}{\delta j^{+}_{2}}{\delta j^{+}_{3}}{j^{+}_{1}}{j^{+}_{2}}{j^{+}_{3}} \;\underset{\beta\geq 0}{\approx}\;\left[\frac{4}{b\pi\delta}\right]^{\frac{1}{4}}\Delta(j^{+}_{1},j^{+}_{2},j^{+}_{3})
\left[\frac{1 + j^{+}_{1} + j^{+}_{2} + j^{+}_{3}}{(1+2j^{+}_{1})(1+2j^{+}_{2})(1+2j^{+}_{3})(1+\delta(j^{+}_{1}+j^{+}_{2}+j^{+}_{3}))}\right]^{\frac{1}{2}} \label{NineJBP}
\end{align}
where, for convenience of notation, we have used the following conventions:
\begin{align}
&b = |\beta| \qquad a = 1 + b = 1 + |\beta| \qquad \delta = 1-b = 1 - |\beta| \\
&\Delta(j^{+}_{1},j^{+}_{2},j^{+}_{3}) = \left[\frac{j^{+}_{1}j^{+}_{2}j^{+}_{3}(1+b(j^{+}_{1}+j^{+}_{2}+j^{+}_{3}))^{-2}}{(j^{+}_{1}+j^{+}_{2}-j^{+}_{3})(j^{+}_{1}-j^{+}_{3}+j^{+}_{3})(-j^{+}_{1}+j^{+}_{2}+j^{+}_{3})(j^{+}_{1}+j^{+}_{2}+j^{+}_{3})}\right]^{\frac{1}{4}}
\end{align}
The EPRL $9J$-symbol (type B) can be easly tabulated compared to the full $9J$-symbol. Upon identifying all three triads we obtain the following asymptotic formulas:
\begin{align}
&\NineJ{bj^{+}}{bj^{+}}{bj^{+}}{j^{+}}{j^{+}}{j^{+}}{aj^{+}}{aj^{+}}{aj^{+}} \;\underset{\beta < 0}{\approx}\;\left[\frac{4a}{3\pi b}\right]^{\frac{1}{4}}\left[\frac{1+3aj^{+}}{\sqrt{j^{+}}(1+3j^{+})(1+3bj^{+})(1+2aj^{+})^{3}}\right]^{\frac{1}{2}}
\;\underset{j\rightarrow +\infty}{\approx}\;j^{-\frac{9}{4}} \label{NineJBM1}\\
&\NineJ{bj^{+}}{bj^{+}}{bj^{+}}{\delta j^{+}}{\delta j^{+}}{\delta j^{+}}{j^{+}}{j^{+}}{j^{+}} \;\underset{\beta\geq 0}{\approx}\;\left[\frac{4}{3\pi\delta b}\right]^{\frac{1}{4}}\left[\frac{1+3j^{+}}{\sqrt{j^{+}}(1+3bj^{+})(1+3\delta j^{+})(1+2j^{+})^{3}}\right]^{\frac{1}{2}}\;
\underset{j\rightarrow +\infty}{\approx}\;j^{-\frac{9}{4}} \label{NineJBP1}
\end{align}
\end{description}
\noindent Last, the following asymptotic formula for $9J$-symbol \cite{Varshalovich:1988ye} is valid when all spins are large but distinct.
\begin{equation}
\Bigg|\NineJ{j_{1}}{j_{2}}{j_{3}}{j_{4}}{j_{5}}{j_{6}}{j_{7}}{j_{8}}{j_{9}}\Bigg|\;\underset{j_{i}\gg 1}{\approx}\frac{L}{\sqrt{(2j_{1}+1)(2j_{2}+1)(2j_{4}+1)(2j_{6}+1)(2j_{8}+1)(2j_{9}+1)}} \qquad L>1 \label{NineJAsy}
\end{equation}
This formula is an approximate upperbound for the modulus of the NineJ symbol in the large-$j$ regime, derived from the complete non-equilateral asymptotic formula for the NineJ symbol which can be found in \cite{Varshalovich:1988ye}, chapter $10$, paragraph $10.7$, eqs. $1-5$.
This formula should be used cautiously since it might not be quite accurate for small spins (e.g. $j\approx 10$). Moreover, it seems to enhance the $9J$-symbol's fall-off rate compared to the above listed formulas.
Hence the non-equilateral asymptotic formulas for the \textit{Type B} NineJ symbol $A.2-A.3$ have been derived by applying the above mentioned eq. $14$ of \cite{Varshalovich:1988ye} to the specific EPRL case and further simplifying the result by using the Stirling approximation for the factorial (to the lowest order). This is possible because upon enforcing the EPRL simplicity constriants a generic NineJ symbol reduces to a NineJ symbol of \textit{type B} to which the equation $14$ applies. Indeed we have:
\begin{align}
&w_{EPRL}(j^{-},j^{+},j,\beta) = \delta_{j^{-}\,|\beta|j^{+}}
\delta_{j\,(1+|\beta|)j^{+}} \quad \beta<0 \qquad 
w_{EPRL}(j^{-},j^{+},j,\beta) = \delta_{j^{-}\,|\beta|j^{+}}
\delta_{j\,(1-|\beta|)j^{+}} \quad \beta\geq 0 \nonumber \\
&\NineJ{j^{-}_{1}}{j^{-}_{2}}{j^{-}_{3}}{j^{+}_{1}}{j^{+}_{2}}{j^{+}_{3}}{j_{1}}{j_{2}}{j_{3}}\; \xrightarrow{EPRL,\,\beta\geq 0}\;
\NineJ{|\beta|j^{+}_{1}}{|\beta|j^{+}_{2}}{|\beta|j^{+}_{3}}{j^{+}_{1}}{j^{+}_{2}}{j^{+}_{3}}{(1-|\beta|)j^{+}_{1}}{(1-|\beta|)j^{+}_{2}}{(1-|\beta|)j^{+}_{3}}
\end{align}
Due to the EPRL constraints each spin in the third row is either the sum or the difference of the other two spins on the same column, depending on the sign of the parameter $\beta$. The equilateral asymptotic formulas for the NineJ symbol $A.6-A.7$ can be obtained in the same way upon enforcing the following additional conditions on the spins: $j^{-}_{1} = j^{-}_{2} = j^{-}_{3} = j^{-}$ and $j^{+}_{1} = j^{+}_{2} = j^{+}_{3} = j^{+}$. 

\subsection[15J-symbol]{$15J$-symbol.} 
The $15J$-symbol (of the first type) can be written either as a single sum of five $6J$-symbols \cite{Varshalovich:1988ye}. We call \textit{tetrad} the set of spins emanating from a given node of the $15J$-symbol graph $\Gamma_{5}$. The $15J$-symbol with one vanishing tetrad obeys the following formula: 
\begin{align}
&\FifteenJ{\MyCol{i_{1}}{0}{j_{1}}}{\MyCol{0}{j_{7}}{i_{2}}}{\MyCol{i_{3}}{j_{8}}{j_{3}}}{\MyCol{j_{4}}{j_{9}}{i_{4}}}{\MyCol{i_{5}}{0}{0}} =  (-1)^{j_{1}+j_{7}+j_{8}+j_{9}}\frac{\delta_{i_{2}i_{5}}\delta_{i_{2}j_{1}}\delta_{i_{5}j_{1}}\delta_{i_{3}j_{7}}\delta_{i_{4}j_{9}}}{(2j_{1}+1)\sqrt{(2j_{7}+1)(2j_{9}+1)}}\SixJ{j_{7}}{j_{3}}{j_{1}}{j_{9}}{j_{4}}{j_{8}} \label{FifteenJD1}
\end{align}
All other formulas can be derived in the same way directly from the definition. 
\subsection{Fusion coefficients and propagators.}
The fusion coefficients are defined as the matrix elements (in the quantum number basis) of a map between the spaces of $4$-valent Spin$(4)$ and SU$(2)$ intertwiners.
\begin{align}
&f^{i,l}_{I}(J_{p}, j_{p},k,\beta) = \sum_{M_{p}m_{p}}(\mathcal{I})^{J_{1}J_{2}J_{3}J_{4}I}_{M_{1}M_{2}M_{3}M_{4}}\bigg[\prod_{p=1}^{4}C^{j^{-}_{p}j^{-}_{p}j_{p}}_{m^{-}_{p}m^{-}_{p}m_{p}}(k)w^{l}(J_{p}, j_{p}, \beta)\bigg](\mathcal{I})^{j_{1}j_{2}j_{3}j_{4}i}_{m_{1}m_{2}m_{3}m_{4}}
\end{align}
Uppercase letters denote Spin$(4)$ representations and intertwiners while lowercase letters label SU$(2)$ data. Upon using the appropriate recoupling identities they can be rewritten as shown below.
\begin{align}
&f^{i,p}_{I}(J_{l}, j_{l},\mathbb{I},\beta) = (-1)^{S}\bigg[\prod_{l=1}^{4}w^{p}(J_{l}, j_{l}, \beta)\bigg]\NineJ{j^{-}_{1}}{i^{-}}{j^{-}_{2}}{j^{+}_{1}}{i^{+}}{j^{+}_{2}}{j_{1}}{i}{j_{2}}\NineJ{j^{-}_{3}}{i^{-}}{j^{-}_{4}}{j^{+}_{3}}{i^{+}}{j^{+}_{4}}{j_{3}}{i}{j_{4}} \nonumber \\
&S = 2(j^{-}_{1} + j^{-}_{2}) + (j^{-}_{3} - j^{+}_{3} + j_{3}) + (j^{-}_{4} - j^{+}_{4} + j_{4}) = 2(j^{-}_{1} + j^{-}_{2}) + S_{3} + S_{4} \label{FCII}
\end{align}
The following identities are particularly useful in the calculation of spin foam amplitudes:
\begin{align}
&\mathcal{K}^{0\,J_{1}J_{2}J_{3}}(l,\beta)\equiv \mathcal{K}^{J_{1}J_{2}J_{3}}(l,\beta) = \sum_{j_{1}j_{2}j_{3}}\prod_{i=1}^{3}d_{j_{i}}w^{l}(J_{i},j_{i},\beta)\NineJ{j^{-}_{1}}{j^{-}_{2}}{j^{-}_{3}}{j^{+}_{1}}{j^{+}_{2}}{j^{+}_{3}}{j_{1}}{j_{2}}{j_{3}}^{2} \label{KEff} \\
&\mathcal{K}^{j^{-}j^{+}}(l,\beta) = \sum_{j_{1}j_{2}j_{3}}
\prod_{i=1}^{3}d_{j_{i}}w^{l}(j^{-},j^{+},j_{i},\beta)
\NineJ{j^{-}}{j^{-}}{j^{-}}{j^{+}}{j^{+}}{j^{+}}{j_{1}}{j_{2}}{j_{3}}^{2} \label{KEffUni}
\end{align}
These formula are completely general. For the (euclidean) EPRL model we have:
\begin{align}
&\mathcal{K}^{J_{1}J_{2}J_{3}}_{\mathrm{EPRL}}(\beta) = \prod_{i=1}^{3}\delta_{j^{-}_{i}|\beta|j^{+}_{i}}d_{(1-\beta)j^{+}_{i}}\NineJ{|\beta|j^{+}_{1}}{|\beta|j^{+}_{2}}{|\beta|j^{+}_{3}}{j^{+}_{1}}{j^{+}_{2}}{j^{+}_{3}}{(1-\beta)j^{+}_{1}}{(1-\beta)j^{+}_{2}}{(1-\beta)j^{+}_{3}}^{2} \label{KEffEPRL} \\
&\mathcal{K}^{j^{-}j^{+}}(\beta<0) = \left[\frac{4a}{3\pi b}\right]^{\frac{1}{2}}\frac{\delta_{j^{-}\,|\beta|j^{+}}(1+3aj^{+})}{\sqrt{j^{+}}(1+3j^{+})(1+3bj^{+})}
\;\underset{j\rightarrow +\infty}{\approx}\;\frac{\delta_{j^{-}\,|\beta|j^{+}}}{(j^{+})^{\frac{3}{2}}} \label{KEffEPRLAsy1} \\
&\mathcal{K}^{j^{-}j^{+}}(\beta\geq 0) = \left[\frac{4}{3\pi\delta b}\right]^{\frac{1}{2}}\delta_{j^{-}\,|\beta|j^{+}}\frac{(1+3j^{+})(1+3\delta j^{+})^{2}}{\sqrt{j^{+}}(1+3bj^{+})(1+2j^{+})^{3}}\;\underset{j\rightarrow +\infty}{\approx}\;\frac{\delta_{j^{-}\,|\beta|j^{+}}}{(j^{+})^{\frac{3}{2}}} \label{KEffEPRLAsy2}
\end{align}
where we have used the identities \ref{KEffUni}, \ref{NineJBM1} and \ref{NineJBP1} together with the EPRL simplicity constraints.

\section[Elements of graph theory.]{Elements of graph theory.} We collect here few elementary graph-theoretic notions often mentioned in the analysis of radiative corrections.
\begin{description}[leftmargin=0pt]
\item {\it Contractible face} - An (internal) face $f$ of a GFT graph 
$\mathcal{G}$ is contractible if it has at least one edge $e$ which is not shared with any other internal face of the same graph.
\item {\it Contractible graph} - A GFT graph $\mathcal{G}$ is contractible if its bulk holonomies $h_{ve}$ can be trivialized in the UV region. A 1PI divergent contractible graph generate a counterterm proportional to a tensor invariant effective interaction (not necessarely a connected bubble.)
\item {\it Tracial graph} - A tracial GFT graph $\mathcal{G}$ is a contractible graph which generates a connected tensor invariant bubble interaction. All melonic graphs are by construction tracial.
\end{description}

\bibliographystyle{JHEP}
\bibliography{Frontiers-GFTrenorm}

\end{document}